\documentclass[11pt]{article}
\usepackage{fullpage}
\usepackage{amsmath}
\usepackage{amssymb}
\usepackage{amsthm}
\usepackage{fancyhdr}
\usepackage{algorithm}
\usepackage{algorithmic}
\usepackage{graphicx}
\usepackage{multirow}
\usepackage{enumitem}                 
\usepackage{bm}
\usepackage{makecell}
\usepackage{scalerel}
\usepackage{dsfont}
\usepackage{caption}
\usepackage[usenames]{color}
\usepackage{cases}
\usepackage{tikz}
\usepackage{comment}
\usepackage[caption=false,font=footnotesize]{subfig}

\allowdisplaybreaks
\newcommand\numberthis{\addtocounter{equation}{1}\tag{\theequation}}
\newtheorem{theorem}{Theorem}[section]
\newtheorem{lemma}[theorem]{Lemma}

\def\ba{\bm{a}}

\def\bx{\bm{x}}
\def\bz{\bm{z}}
\def\bu{\bm{u}}
\def\bv{\bm{v}}
\def\bw{\bm{w}}
\def\by{\bm{y}}
\def\bq{\bm{q}}

\def\N{\mathcal{N}}
\def\R{\mathbb{R}}
\def\S{\mathcal{S}}
\def\MR{\mathcal{R}}
\def\BA{\bm{A}}
\def\BB{\bm{B}}
\def\A{\mathcal{A}}

\def\BX{\bm{X}}
\def\BZ{\bm{Z}}
\def\BI{\bm{I}}
\def\TM{\mathbf{I}}
\def\grad{\nabla}
\def\hessian{\nabla^2}

\def\lb{\left(}
\def\rb{\right)}
\def\ln{\left\|}
\def\rn{\right\|}
\def\lab{\left|}
\def\rab{\right|}
\def\lcb{\left\{}
\def\rcb{\right\}}
\def\lsb{\left[}
\def\rsb{\right]}

\DeclareMathOperator*{\dist}{\mbox{\normalfont dist}}
\DeclareMathOperator*{\rank}{\mbox{\normalfont rank}}

\newcommand{\mean}[1] {\mathbb{E}{\lsb#1\rsb}}

\def\repE{\mean{\bm{\cdot}}}
\newcommand{\dsone}[1]{\mathds{1}_{\lcb#1\rcb}}

\title{Towards the optimal construction of a loss function without spurious local minima for solving quadratic equations}
\author{Zhenzhen Li\thanks{Department of Mathematics, Hong Kong University of Science and Technology, Clear Water Bay, Kowloon, Hong Kong SAR, China; Email: zlice@ust.hk.}
\and 
Jian-Feng Cai\thanks{Department of Mathematics, Hong Kong University of Science and Technology, Clear Water Bay, Kowloon, Hong Kong SAR, China; Email: jfcai@ust.hk.}
\and Ke Wei\thanks{School of Data Science, Fudan University, Shanghai, China; Email: kewei@fudan.edu.cn.}}
\begin{document}
\maketitle

\begin{abstract}
The problem of finding a vector $\bx$ which obeys a set of quadratic equations $|\ba_k^\top\bx|^2=y_k$,  $k=1,\cdots,m$, plays an important role in many applications. In this paper we consider the case when both $\bx$ and $\ba_k$ are real-valued vectors of length $n$. A new loss function is constructed for  this problem, which combines  the smooth quadratic loss function with an activation function. Under the Gaussian measurement model, we establish that with high probability the target solution $\bx$ is the unique  local minimizer (up to a global phase factor) of the new loss function provided $m\gtrsim n$. Moreover, the loss function always has a negative directional curvature around its saddle points.
\end{abstract}

\section{Introduction}\label{sec:introduction}
Many applications in science and engineering, such as X-ray crystallography \cite{Ha93a},  diffraction and array imaging \cite{Buetal07}, and electron microscopy \cite{Mietal08}, are essentially about solving systems of quadratic equations. This paper concerns a real-valued case of the problem. The goal is to find a vector $\bx\in\R^n$ which can solve $m$ quadratic equations of the form
\begin{align*}
y_k=|\ba_k^\top\bx|^2,\quad k=1,\cdots,m,\numberthis\label{eq:problem}
\end{align*}
where  $\by=\begin{bmatrix}y_1,\cdots,y_m
\end{bmatrix}^\top\in\R_+^{m}$ and { $\{\bm{a}_k\in\mathbb{R}^n\}_{k=1}^m$} are known.  Despite the seeming simplicity of \eqref{eq:problem}, solving this problem is computationally intractable. Indeed, a special instance of  \eqref{eq:problem} is the NP-hard stone problem \cite{CC:CPAM:17}.

The problem of recovering a vector from a set of quadratic measurements, especially from the Fourier type measurements, has long been studied. Moreover, it has received intensive investigations over the past few years largely due to its connection with low rank matrix recovery. Even though the corresponding low rank matrix recovery problem is  still nonconvex and computationally intractable, 
we can approximate it  by its nearest convex relaxation, leading to a convex formulation known as PhaseLift. 
Performance guarantee of PhaseLift has been established in \cite{CSV:CPAM:13,CL:FCM:14,phaselift4,phaselift5} under different measurement models, showing that successful recovery can be achieved when the number of equations is (nearly) proportional to the number of unknowns. There are also other convex relaxation methods for solving systems of quadratic equations; see for example \cite{phasecut,phasemax1,phasemax2,phasemax3}.

Though convex approximations usually come with recovery guarantees,  they are not computationally desirable for { large-scale problems}. In contrast, many simple nonconvex algorithms are able to solve \eqref{eq:problem} both accurately and efficiently. Among them are a family of algorithms with optimal or near-optimal provable guarantees, including  alternating projections and its resampled variant \cite{alt_min_pr,Waldspurger16}, Kaczmarz methods \cite{phase_kacz02,phase_kacz03}, and those algorithms which propose to compute the solution of \eqref{eq:problem} by minimizing certain nonconvex loss functions { \cite{CLS:TIT:15,CC:CPAM:17,WGE:TIT:18,CaiWeiphase,ReshapedWF,RAF2018}}. Specifically,  a gradient descent algorithm known as Wirtinger Flow has been developed in 
\cite{CLS:TIT:15} based on the following smooth quadratic loss function 
\begin{align*}
{ \tilde{f}(\bz)}=\frac{1}{2m}\sum_{k=1}^m\lb ({ \ba_k^\top}\bz)^2-y_k\rb^2.\numberthis\label{eq:loss1}
\end{align*}
In \cite{WGE:TIT:18,ReshapedWF}, { gradient descent} algorithms were developed based on a loss function similar to \eqref{eq:loss1} but with $(\ba_k^\top\bz)^2-y_k$ replaced by $|\ba_k^\top\bz|-\sqrt{y_k}$, while a Poisson loss function is adopted in \cite{CC:CPAM:17}.

Theoretical guarantees of the aforementioned algorithms typically require that the initial guess is sufficiently close to the true solution. However, numerical simulations show that these algorithms can often achieve successful recovery even with random initialization.  To understand this empirical success, Sun et al. \cite{SQW:FCM:18} investigated the global geometry of the loss function in \eqref{eq:loss1}. It has been shown that under the Gaussian measurement model { $\tilde{f}(\bz)$} does not have any spurious local minima provided\footnote{The notation $m\gtrsim { g(n)}$ means that there exists an absolute constant $C>0$ such that $m\ge C\cdot{  g(n)}$. {  It is worth noting that $m\gtrsim n\log^3n$ in \cite{SQW:FCM:18} is a sufficient condition for the well-behaved landscape of $\tilde{f}(\bz)$, so this does not mean that  a spurious { minimum} of $\tilde{f}(\bz)$ exists when $n\lesssim m\lesssim n\log^3n$.}} $m\gtrsim n\log^3n$. Putting it in another way, under this sampling condition, the target signal $\bx$ is { the only  minimizer} of { $\tilde{f}(\bz)$} up to a global phase factor. Moreover,  { $\tilde{f}(\bz)$} possesses a negative directional curvature around each saddle point. Thus, algorithms that can avoid  saddle points and converge to a local minimizer  are bound to find { a global minimizer}; see for example \cite{LSJR:COLT:16}.  Our work  follows this line of research and attempts to construct a loss function with $\bx$ being { the only minimizer} up to a global phase factor when $m\gtrsim n$.
That is, we want to construct a loss function without spurious local minima for \eqref{eq:problem} conditioned on the optimal sampling complexity. 

In recent years, there has been a surge of interest in nonconvex optimization for problems arising from signal processing and machine learning; solving systems of quadratic equations is one of them. For more general low rank matrix recovery, a variety of nonconvex algorithms have been developed and analyzed, including those based on matrix factorization \cite{TBSSR:ICML:16,ZL:ArXiv:16} and those based on the embedded manifold of low rank matrices \cite{WCCL:SIMAX:16,WCCL:ArXiv:16}. The reader can refer to the review paper \cite{CaiWeiReview} for more details. Geometric landscape of related loss functions for low rank matrix recovery has been investigated in \cite{GJZ:ICML:17,GLM:NIPS:16,LiZhTa16,BNS16a,PKCS:ArXiv:16}.  Similar results have also been established for nonconvex formulations of other problems, for example blind deconvolution \cite{ZLKC:CVPR:17}, dictionary learning \cite{SQW:TIT:17:I,SQW:TIT:17:II}, tensor completion \cite{AnGeJa,GeMa:170a}, phase synchronization \cite{BourmalSyn16, LYSo170a,BoVoBa:ArXiv:18}, and deep neural networks \cite{WeBaBruna:ArXiv:18,YunSraJad:ArXiv:18,SoudryCarmon:ArXiv:16,Kawaguchi:ArXiv:16}.
\subsection{Motivation and main result}
As stated previously, a few of the  algorithms for solving Gaussian random systems of quadratic equations  are able to achieve successful recovery with high probability
provided  $m\gtrsim n$, including TWF \cite{CC:CPAM:17}, TAF \cite{WGE:TIT:18} and TRGrad \cite{CaiWeiphase}, just to name a few. In addition, it is also known that a unique solution (up to a global phase factor) of \eqref{eq:problem} can be determined from $m\geq 2n-1$ generic measurements  for the real problem or from $m\geq 4n-4$ generic measurements for the complex problem \cite{BaCaEd06a,CoEdHeVi2013a}. Thus, it is interesting to see whether 
there exists a loss function for solving random systems of quadratic equations which does not have any spurious local minima when $m\gtrsim n$, in contrast to $m\gtrsim n\log^3 n$ for \eqref{eq:loss1} as is established in \cite{SQW:FCM:18}. To the best of our knowledge, this question has not been explored yet. In our work, we will give an affirmative answer for the real-valued problem.

\begin{figure}[ht!]
\centering
	\includegraphics[width=0.4\textwidth]{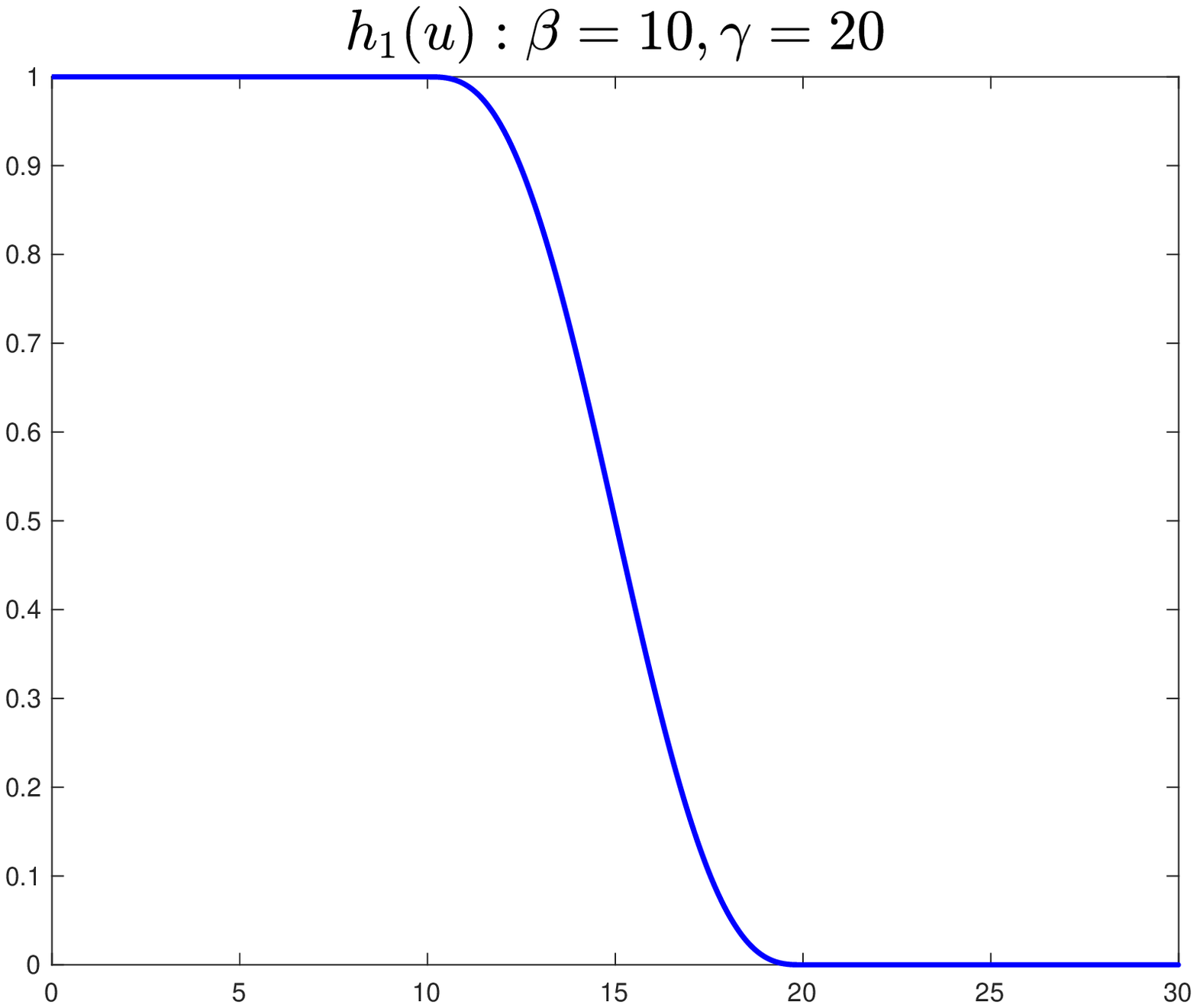}\hfil
	\includegraphics[width=0.4\textwidth]{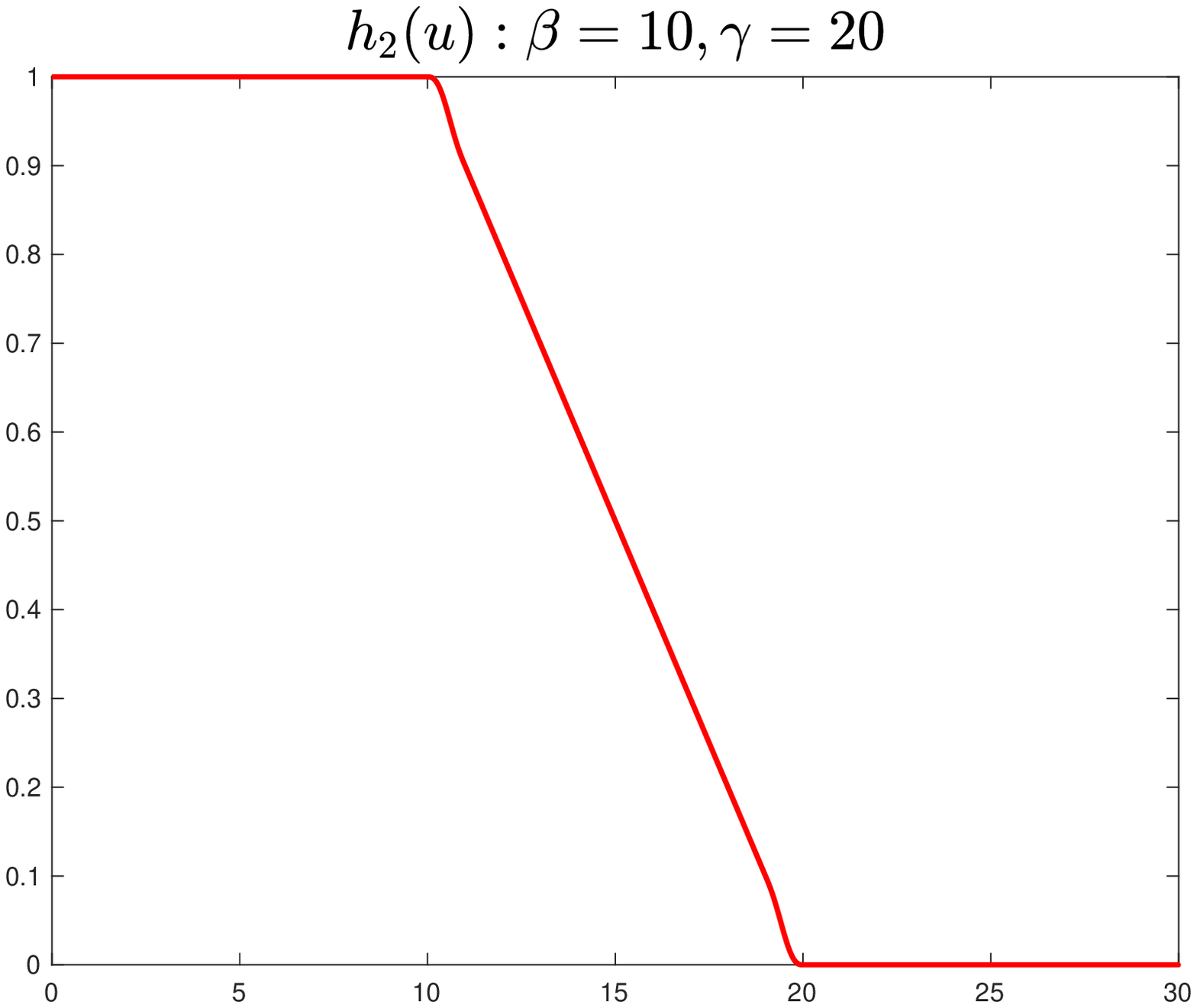}
\caption{Two examples of activation functions: $h_1(u)$ (left) and $h_2(u)$ (right).}
\label{fig1}
\end{figure}

We construct the new loss function $f(\bz)$ by coupling  \eqref{eq:loss1} with an activation function $h(u)$, 
\begin{align*}
f(\bz)=\frac{1}{2m}\sum_{k=1}^m\lb(\ba_k^\top\bz)^2-(\ba_k^\top\bx)^2\rb^2h\lb\frac{|\ba_k^\top\bz|^2}{\|\bz\|^2}\rb h\lb\frac{m|\ba_k^\top\bx|^2}{\|\by\|_1}\rb,\numberthis\label{eq:loss2}
\end{align*}
where the activation function $h(u)$ satisfies
\begin{align*}
\begin{cases}
h(u) = 1 & \mbox{if }0\leq u\leq \beta,\\
h(u)\in[0,1] &\mbox{if }u\in(\beta,\gamma),\\
h(u)=0&\mbox{if }u\geq \gamma
\end{cases}
\quad\mbox{and}\quad
|h'(u)|,~|h''(u)| \mbox{ exist and are bounded }
\end{align*}
for two predetermined universal parameters { $1<\beta<\gamma$} that are sufficiently large.  {  As can be seen later, the activation function has been introduced to control the gradient of the function so that overshooting can be avoided.}

For simplicity, we assume $\gamma=C\cdot\beta$ for some absolute constant $C>1$. Note that 
the bounds of $|h'(u)|$ and $|h''(u)| $ { rely on} the parameters $\beta$ and $\gamma$. Two examples of $h(u)$ are  
\begin{align*}
h_1(u) = \begin{cases}
1 & 0\leq u\leq \beta\\
-6\lb\frac{u-\beta}{\gamma-\beta}\rb^5+15\lb\frac{u-\beta}{\gamma-\beta}\rb^4-10\lb\frac{u-\beta}{\gamma-\beta}\rb^3+1 & u\in (\beta,\gamma)\\
0 & u \geq \gamma.
\end{cases}
\end{align*}
and
\begin{align*}
h_2(u) = \begin{cases}
1 & 0\leq u\leq \beta\\
-30000\lb\frac{u-\beta}{\gamma-\beta}\rb^5+8000\lb\frac{u-\beta}{\gamma-\beta}\rb^4-600\lb\frac{u-\beta}{\gamma-\beta}\rb^3+1 & 0<\frac{u-\beta}{\gamma-\beta}<0.1 \\ 1-\frac{u-\beta}{\gamma-\beta}&
0.1\leq \frac{u-\beta}{\gamma-\beta}\leq 0.9\\
-30000\lb\frac{u-\beta}{\gamma-\beta}-1\rb^5-8000\lb\frac{u-\beta}{\gamma-\beta}-1\rb^4-600\lb\frac{u-\beta}{\gamma-\beta}-1\rb^3&0.9<\frac{u-\beta}{\gamma-\beta}<1\\
0 & u \geq \gamma.
\end{cases}
\end{align*}
See Figure~\ref{fig1} for a graphical illustration of $h_1(u)$ and $h_2(u)$ when $\beta=10$ and $\gamma=2\beta$.  {  The smoothness of $h_1(u)$ and $h_2(u)$ can be verified directly. Indeed, a direct calculation  yields that
\begin{align*}
\begin{cases}
h_1'(\beta-)=h_1'(\beta+)=0,~h_1''(\beta-)=h_1''(\beta+)=0,\\
h_1'(\gamma-)=h_1'(\gamma+)=0,~h_1''(\gamma-)=h_1''(\gamma+)=0,
\end{cases}
\end{align*}
and
\begin{align*}
\begin{cases}
h_2'(\beta-)=h_2'(\beta+)=0,~h_2''(\beta-)=h_2''(\beta+)=0,\\
h_2'((0.1(\gamma-\beta)+\beta)-)=h_2'((0.1(\gamma-\beta)+\beta)+)=-\frac{1}{\gamma-\beta},~h_2''((0.1(\gamma-\beta)+\beta)-)=h_2''((0.1(\gamma-\beta)+\beta)+)=0,\\
h_2''((0.9(\gamma-\beta)+\beta)-)=h_2'((0.9(\gamma-\beta)+\beta)+)=-\frac{1}{\gamma-\beta},~h_2''((0.9(\gamma-\beta)+\beta)-)=h_2'((0.9(\gamma-\beta)+\beta)+)=0,\\
h_2'(\gamma-)=h_2'(\gamma+)=0,~h_2''(\gamma-)=h_2''(\gamma+)=0.
\end{cases}
\end{align*}
}

\begin{figure}[!t]
\centering
\includegraphics[width=.35\textwidth]{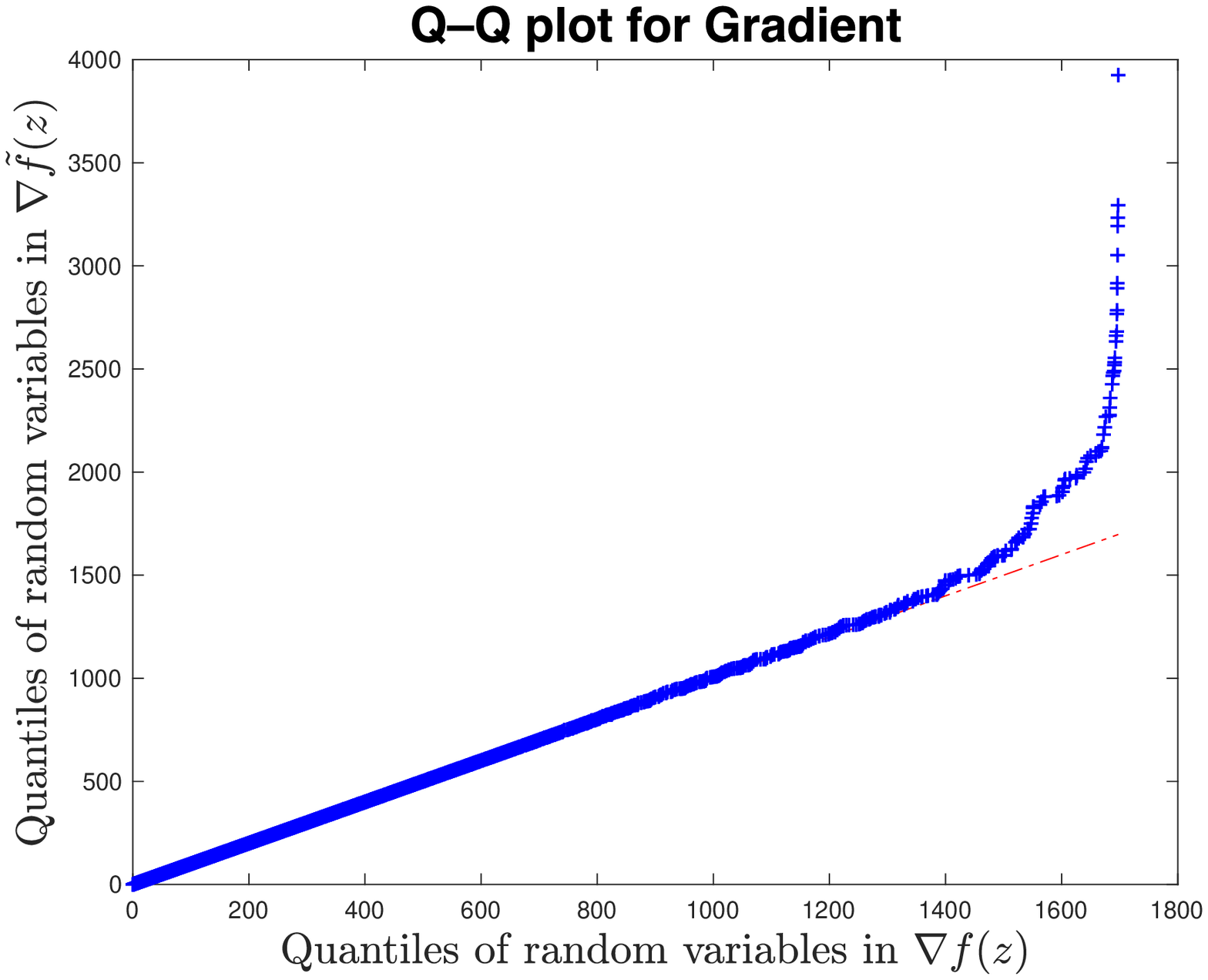}
\hfil
\includegraphics[width=.35\textwidth]{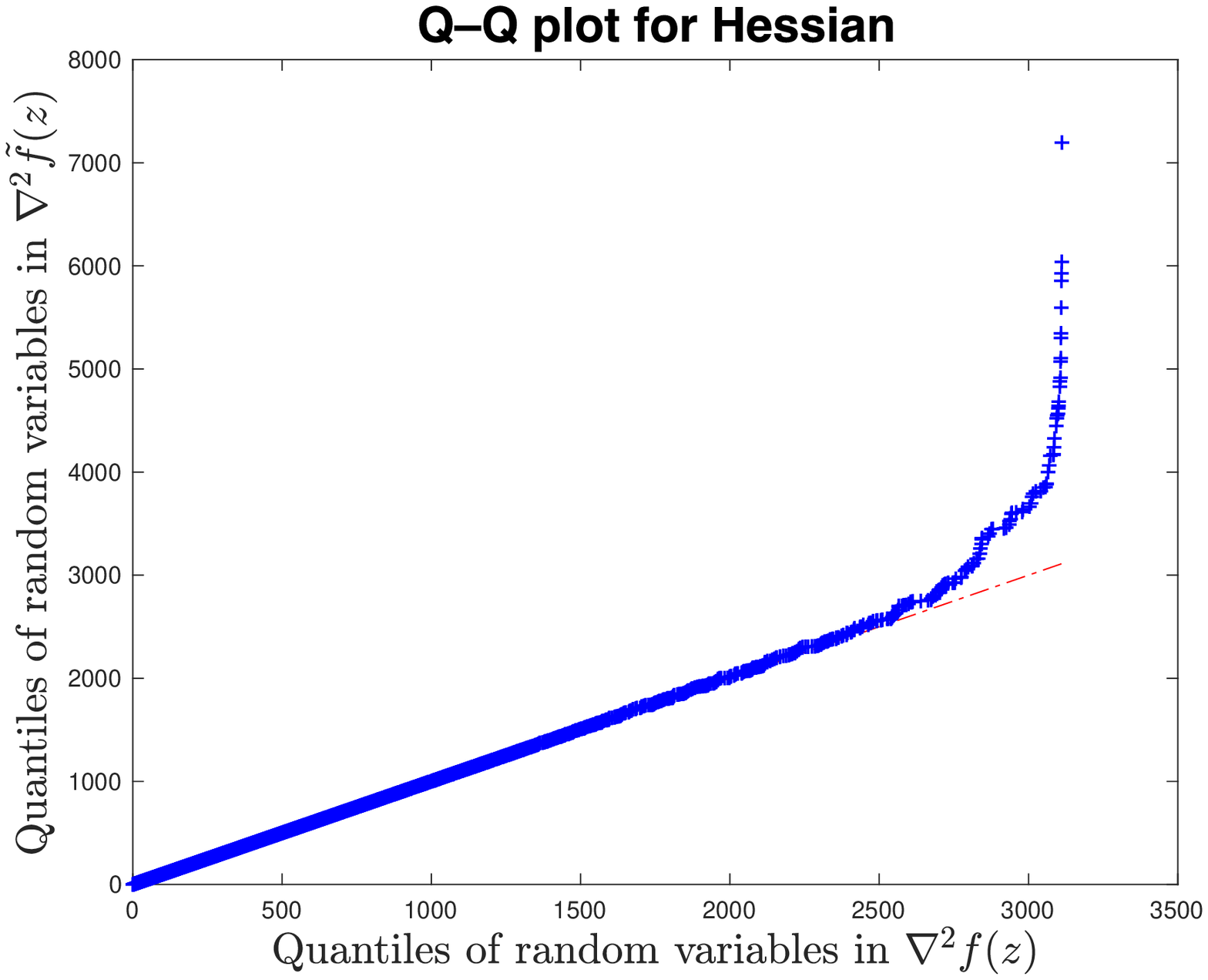}
\caption{ Q-Q plots for the random variables in the gradients (left) and Hessians (right) of ${f}(\bz)$ and $\tilde{f}(\bz)$ when $n=1$. In this simulation, $x=1$, $z=2$, and a total of $10^5$ (i.e., $m=10^5$) standard Gaussian random variables are independently generated. Then we compute the first and second derivatives of each term in \eqref{eq:loss1} and \eqref{eq:loss2}, respectively. The activation function $h_1(u)$ with $\beta=10$ and $\gamma=20$ is used in $f(\bz)$. }\label{fig:QQ}
\end{figure}

{ The introduction of the activation function makes the gradient and Hessian of $f(\bz)$ more complicated, for example the Hessian of $f(\bz)$ has 12 terms. Thus we  postpone the calculations of $\nabla f(\bz)$ and $\nabla^2 f(\bz)$ to Appendix~\ref{app:gradHess}. Despite this, 
the activation function is able to circumvent the effect of the { fourth power} of Gaussian random variables which is heavy-tailed. To demonstrate this effect, we consider the case when $n=1$ (i.e., both $z$ and $x$ are scalars) and then use the Q-Q plot to compare the random variables in the expressions of the gradients (indeed derivative since $n=1$) of $\nabla \tilde{f}(\bz)$ and $\nabla f(\bz)$, as well as the random  variables in the expressions of the Hessians (indeed second derivative since $n=1$) of $\nabla^2 \tilde{f}(\bz)$ and $\nabla^2 f(\bz)$. The plots are presented in Figure~\ref{fig:QQ}, from which we can clearly see that the random variables without the activation function are more heavy-tailed.  } 

Assuming that $\ba_k$, $k=1,\cdots,m$ are independent Gaussian vectors: $\ba_k\sim\N(0,\BI_n)$, our main result for $f(\bz)$ is stated { as follows.}

\begin{theorem}[Main result] \label{thm:main}
With probability exceeding\footnote{{  Here $\Omega(m)$ is a value which is greater than $C\cdot m$ for some numerical constant $C>0$.}} $1-e^{-\Omega(m)}$, the function $f(\bz)$ { defined in \eqref{eq:loss2}} with sufficiently large $1<\beta<\gamma$ does not have any spurious local minima provided $m\gtrsim n$. Moreover,  at each saddle point $f(\bz)$ has a negative directional { curvature.} 
\end{theorem}

{ We would like to note that $\beta$ and $\gamma$ in Theorem~\ref{thm:main} are two absolute positive constants whose values are fixed, and the constant hidden in $m\gtrsim n$ relies on $\beta$ and $\gamma$.
}

\subsection{Numerical Illustration}
\begin{figure*}[!t]
\centering
\subfloat[]{\includegraphics[width=.35\textwidth]{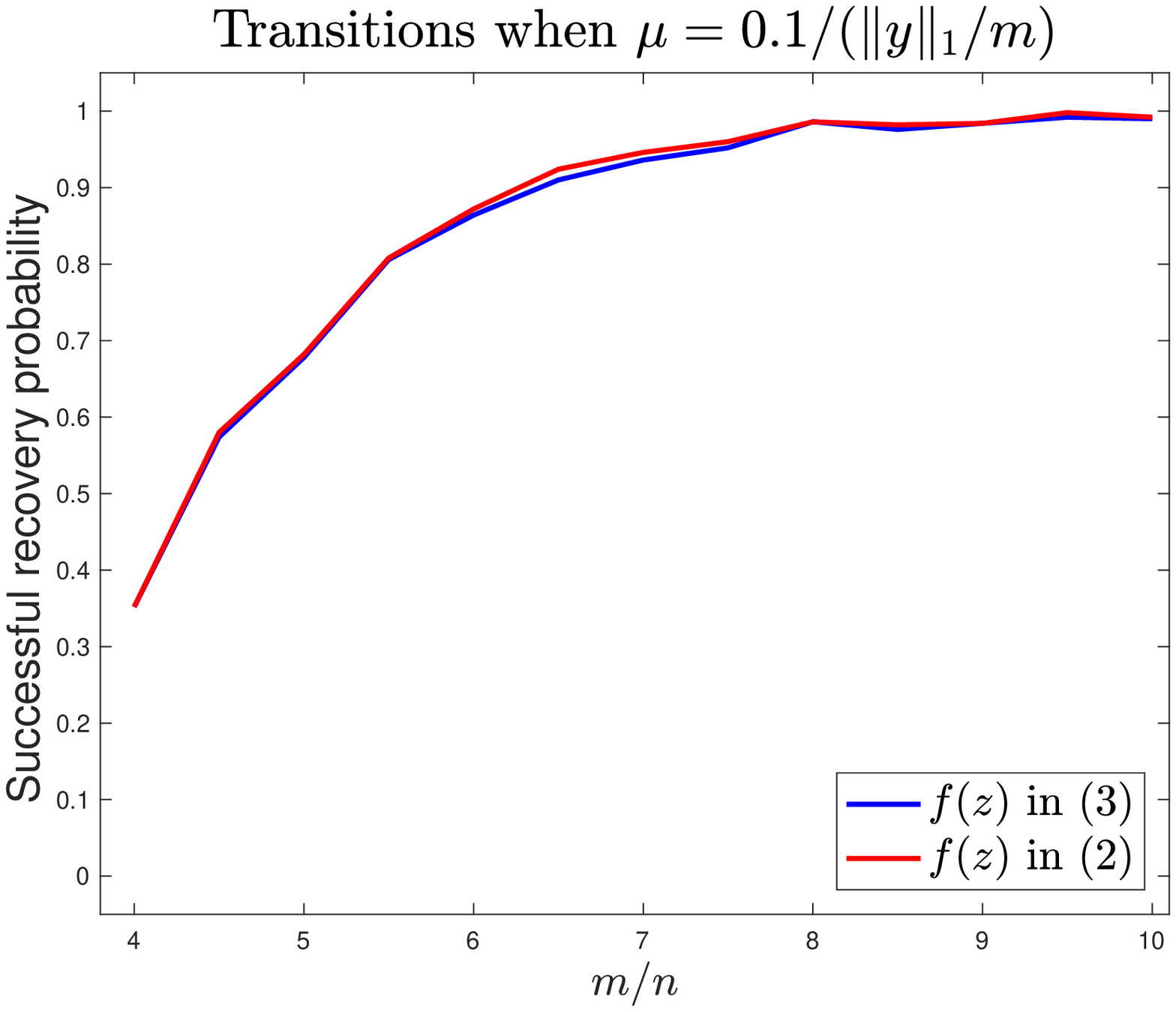}
\label{mu01}}
\hfil
\subfloat[]{\includegraphics[width=.35\textwidth]{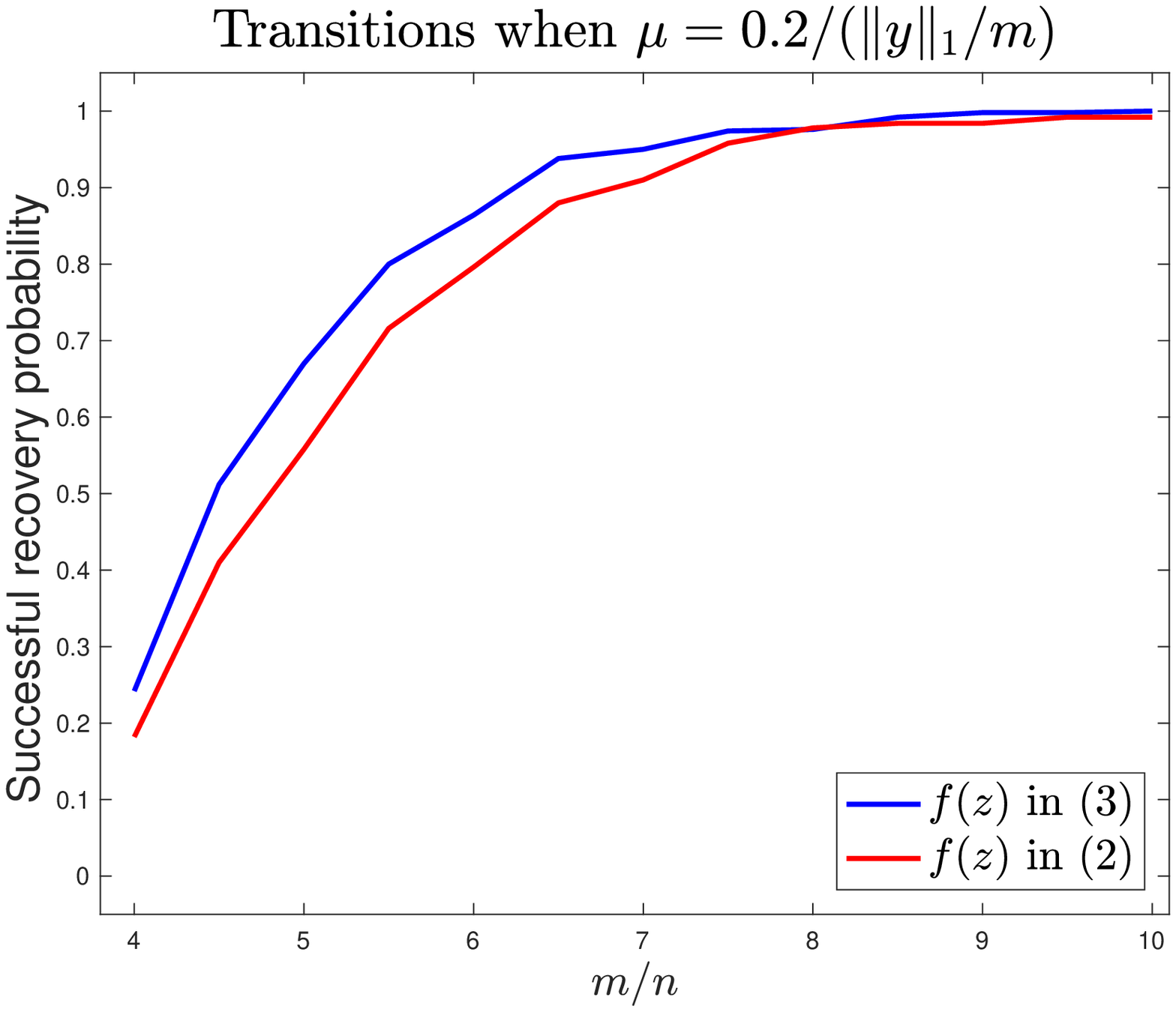}
\label{mu02}}\\
\subfloat[]{\includegraphics[width=.35\textwidth]{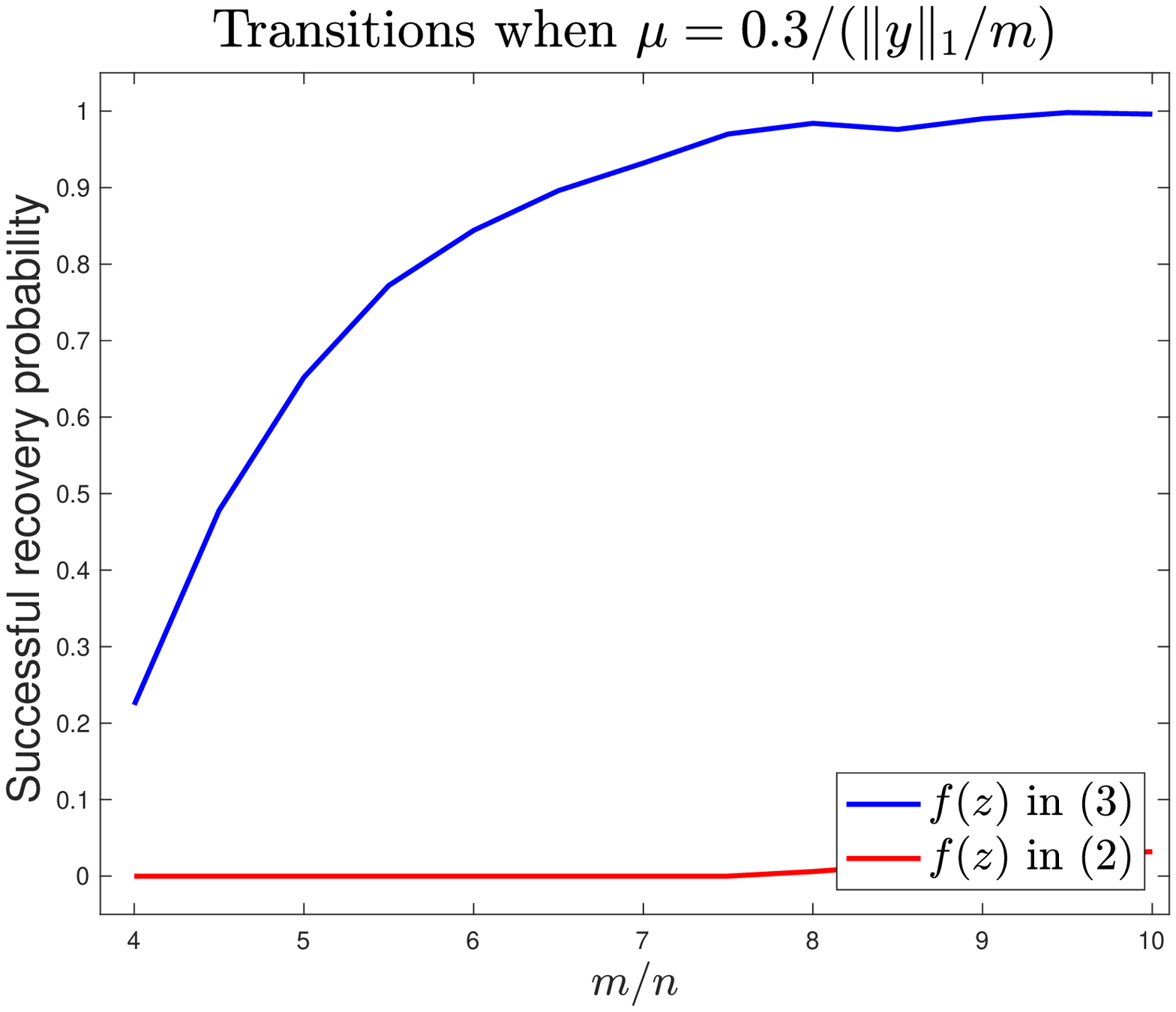}
\label{mu03}}
\hfil
\subfloat[]{\includegraphics[width=.35\textwidth]{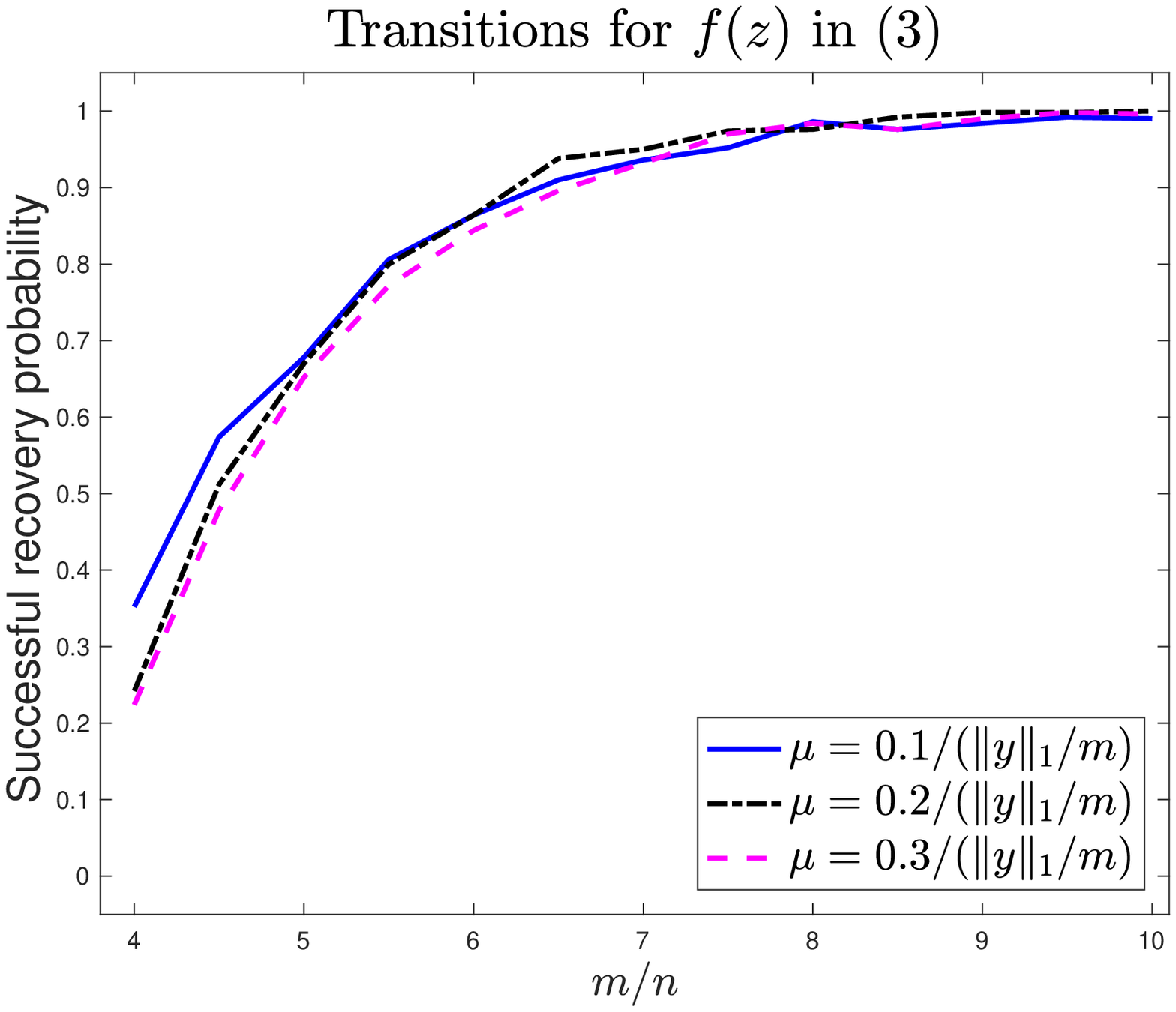}
\label{muall}}
\caption{Recovery performance of gradient descent for functions in \eqref{eq:loss1} and \eqref{eq:loss2}.}
\end{figure*}
In numerical simulations, a direct examination of the geometric landscape of a loss function seems to be out of reach. Instead, we investigate the performance of the gradient descent iteration 
\begin{align*}
{ \bz_{l+1} = \bz_l - \mu\nabla \tilde{f}(\bz_l)\quad\mbox{and}}\quad\bz_{l+1} = \bz_l - \mu\nabla f(\bz_l)
\end{align*}
with three different stepsizes $\mu\in\{0.1,0.2,0.3\}/(\|\by\|_1/m)$ when minimizing the loss functions defined in \eqref{eq:loss1} and \eqref{eq:loss2}, respectively. We use $h_2(u)$ with $\gamma=1.5\beta$ for  the loss function in \eqref{eq:loss2}. Different values of $\beta$ are adopted   for different stepsizes, namely, $\beta=20$ when $\mu=0.1/(\|\by\|_1/m)$, $\beta=10$ when $\mu=0.2/(\|\by\|_1/m)$, and $\beta=5$ when $\mu=0.3/(\|\by\|_1/m)$. Roughly speaking, a more stringent activation condition is imposed for the larger stepsize.

Numerical tests are conducted for fixed $n=128$ and $m/n$ increasing from $4$ to $10$ by $0.5$. For each fixed pair of $(n,m)$, $500$ problem instances on randomly generated $\ba_k\sim\N(0,\BI_n)$  and $\bx\sim\N(0,\BI_n)$ are tested. The initial guess for the gradient descent iteration 
is generated randomly and independently according to the standard Gaussian distribution. We consider the algorithm to have successfully reconstructed a test signal if it returns an estimate with the relative reconstruction error { $\dist(\bz_l,\bx)/\|\bx\|$} being { less than  or equal to} $10^{-3}$ under the distance defined by
\begin{align*}
\dist(\bz,\bx){ =} \min\{\|\bz-\bx\|,\|\bz+\bx\|\}.
\end{align*}
The plots of the successful recovery probability against the sampling ratio for the three different stepsizes are presented in Figures~\ref{mu01} -- \ref{mu03}. We can see that, when 
$\mu=0.1/(\|\by\|_1/m)$, the transition curves of the gradient iterations based on the two different loss functions are nearly indistinguishable. However, the advantage of our  loss function over the one without the activation function becomes more significant as $\mu$ increases. In particular, when $\mu=0.3/(\|\by\|_1/m)$, the gradient iteration based on the new loss function with proper ($\beta$, $\gamma$) can achieve more than $80$\% successful recovery when $m\geq 6n$, whereas the gradient descent iteration based on the other loss function can hardly succeed even when $m=10n$. {  A close look at the simulation results reveals that the gradient descent method for the vanilla $\ell_2$ loss function $\tilde{f}(\bz)$ can either diverge or converge to a local minimizer when $\mu=0.3/(\|\by\|_1/m)$.  In contrast, the new loss function can still succeed for larger stepsizes potentially because the activation function can regularize each component of the gradient so that those components that go wild can be  avoided. }

We also put the recovery transitions corresponding to  the new loss function but with different values of ($\mu$, $\beta$, $\gamma$) in the same plot; see Figure~\ref{muall}. 
Competitive performance of the gradient descent iterations corresponding to different triples of ($\mu$, $\beta$, $\gamma$) can be observed when $m\gtrsim 5n$.
This suggests that similar recovery performance can be achieved by trading off appropriately between the stepsize and the parameters in the loss function.
\subsection{Organization and notation}
The rest of this paper is organized as follows. The geometric landscape of the new loss function is presented in Section~\ref{sec:results}. Section~\ref{sec:main_proofs} contains the detailed justification, with the proofs for the technical lemmas being presented in Section~\ref{sec:tech_proofs}. We conclude this paper with potential future directions in Section~\ref{sec:conclusion}.

Following the notation above we use bold face lowercase letters to denote { column} vectors and use normal font letters with subindices for their entries. In particular, we fix $\bx$ as the underlying vector to be reconstructed. The $\ell_1$-norm and $\ell_2$-norm of a vector $\bz$ are denoted by $\|\bz\|_1$ and $\|\bz\|$, respectively. { Given two vectors $\bz$ and $\bx$, their distance, denoted $\dist(\bz,\bx)$, is defined by
\begin{align*}
\dist(\bz,\bx){=} \min\{\|\bz-\bx\|,\|\bz+\bx\|\}.
\end{align*}}{For a given matrix $\BA$, we use $\|\BA\|$ to denote the matrix operator norm which is defined by
\begin{align*}
\|\BA\| = \sup_{\|\bz\|=1}\|\BA\bz\|.
\end{align*}
When $\BA$ is symmetric we also have
\begin{align*}
\|\BA\|=\sup_{\|\bz\|=1}|\bz^\top\BA\bz|.
\end{align*}
For two symmetric positive semidefinite matrices $\BA$ and $\BB$, if $\BA\preceq \BB$ then $\|\BA\|\leq \|\BB\|.$}

Recall that the notation $m\gtrsim { g(n)}$ means that there exists an absolute constant $C>0$ such that $m\ge C\cdot { g(n)}$. Similarly,  the notation $m\lesssim { g(n)}$ means that there exists an absolute constant $C>0$ such that $m\leq C\cdot { g(n)}$. Throughout the paper, $C$ denotes an absolute constant whose value may change from line to line. {  In addition, $!!$ means double factorial; that is $n!!=n(n-2)(n-4)\cdots$.}
\section{Geometric landscape of the new function}\label{sec:results}
\begin{figure}
\centering
\includegraphics[width=0.75\textwidth]{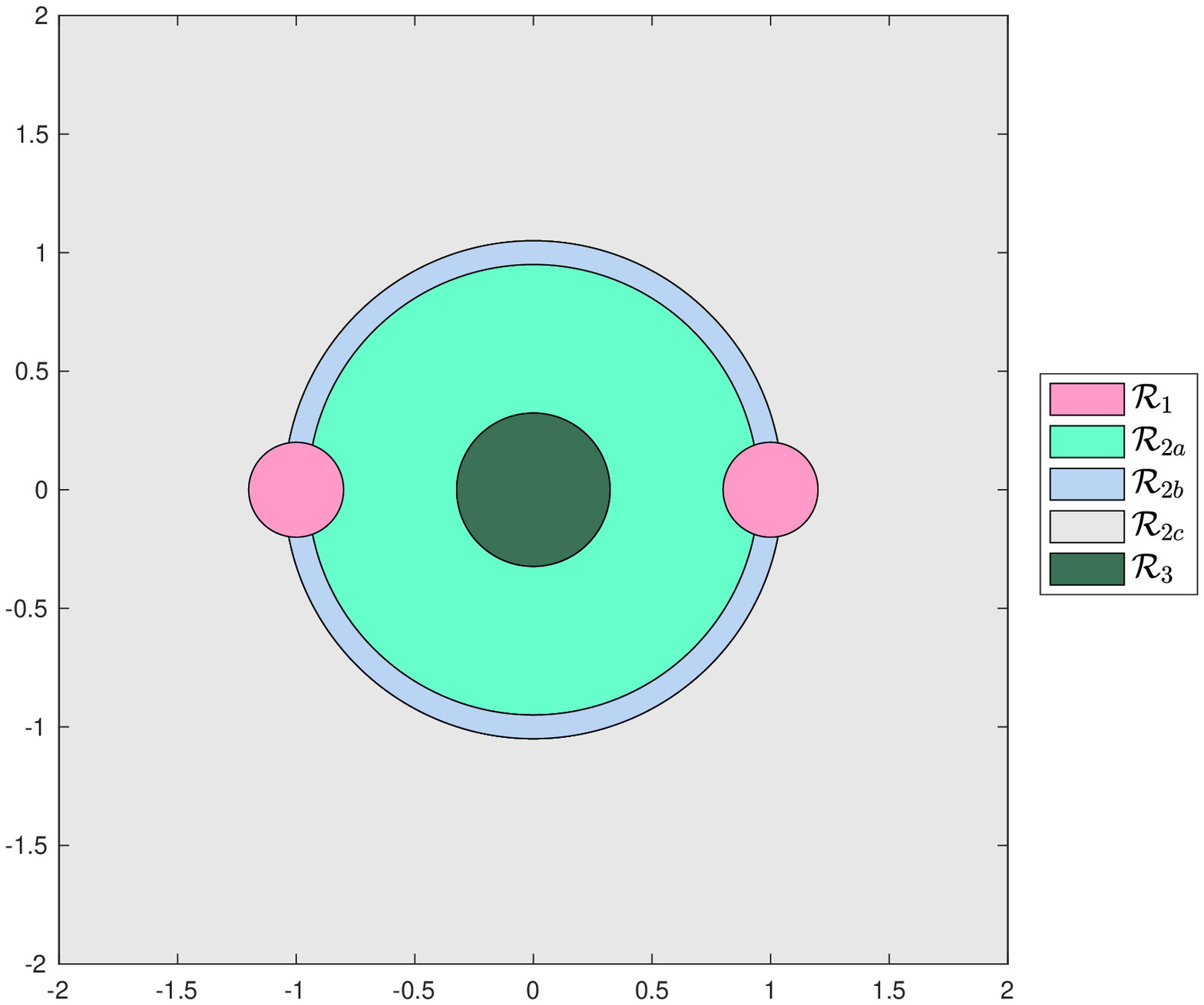}
\caption{  Partition of $\R^2$: $\bx=[\pm1,0]^\top$.}\label{fig2}
\end{figure}

In this section we present the detailed geometric landscape of $f(\bz)$. Differing from the partition in \cite{SQW:FCM:18}, we decompose $\R^n$ into five non-overlapping regions (see Figure~\ref{fig2}):
\begin{itemize}
\item $\MR_1:= \lcb\bz:~\dist(\bz,\bx)\leq \frac{1}{5}\|\bx\|\rcb$, 
\item $\MR_{2a}:=\{\bz:~\frac{1}{3}-\delta<\frac{\|\bz\|^2}{\|\bx\|^2}<\frac{99}{100}\mbox{ and }\dist(\bz,\bx)>\frac{1}{5}\|\bx\| \}$,
\item $\MR_{2b}:=\{\bz:~\frac{99}{100}\leq \frac{\|\bz\|^2}{\|\bx\|^2}\leq\frac{101}{100}\mbox{ and }\dist(\bz,\bx)>\frac{1}{5}\|\bx\|\}$,
\item $\MR_{2c}:=\{\bz:~\frac{\|\bz\|^2}{\|\bx\|^2}>\frac{101}{100}\mbox{ and }\dist(\bz,\bx)>\frac{1}{5}\|\bx\|\}$,
\item $\MR_{3}:=\{ \bz:~0<\frac{\|\bz\|^2}{\|\bx\|^2}\leq \frac{1}{3}-\delta\}$,
\end{itemize}
where  $\delta$ is a fixed constant in $(0,\frac{1}{100}]$. The properties of $f(\bz)$ over these five regions are summarized in the following five theorems.

\begin{theorem}\label{thm:R1}
With probability at least $1-e^{-\Omega(m)}$,
\begin{align*}
\lambda_{\min}\lb\nabla^2f(\bz)\rb\ge \frac{1}{25}\|\bx\|^2
\end{align*}
holds uniformly for all   $ \bz\in\MR_1$ provided  $m\gtrsim n$.
\end{theorem}

\begin{theorem}\label{thm:R2a}
With probability at least $1-e^{-\Omega(m)}$, all critical points in $\MR_{2a}$ must exist in the subregion defined by
\begin{align*}
\frac{1}{3}-\delta<\frac{\|\bz\|^2}{\|\bx\|^2}<\frac{1}{3}+\delta\quad\mbox{and}\quad
|\bz^\top\bx|<\delta\|\bx\|^2\numberthis\label{eq:SubofR2a}
\end{align*}
provided $m\gtrsim n$.  Moreover, with probability exceeding $1-e^{-\Omega(m)}$,

\begin{align*}
\bx^\top\hessian f(\bz)\bx \leq -3\|\bx\|^4\quad\mbox{and}\quad\bz^\top\hessian f(\bz)\bz\geq \|\bx\|^4.
\end{align*}
hold uniformly for all $\bz$ in { the subregion} \eqref{eq:SubofR2a} provided $m\gtrsim n$. \end{theorem}
\begin{theorem}\label{thm:R2b}
With probability at least $1-e^{-\Omega(m)}$,
\begin{align*}
\bz^\top\nabla f(\bz)\geq \frac{9}{1000}\|\bx\|^4
\end{align*}
holds uniformly for all $\bz\in\MR_{2b}$  provided $m\gtrsim n$.
\end{theorem}
\begin{theorem}\label{thm:R2c}
With probability at least $1-e^{-\Omega(m)}$,
\begin{align*}
\bz^\top\nabla f(\bz)\geq \frac{49}{1000}{\|\bz\|^4}
\end{align*}
holds uniformly for all $\bz\in\MR_{2c}$ provided $m\gtrsim n$.
\end{theorem}
\begin{theorem}\label{thm:R3}
With probability at least $1-e^{-\Omega(m)}$,
\begin{align*}
\bz^\top\nabla f(\bz) \leq -5\delta\|\bz\|^2\|\bx\|^2
\end{align*}
holds  uniformly for all $\bz\in\MR_3$ provided $m\gtrsim n$.
\end{theorem}
The proofs of the above theorems are deferred to Section~\ref{sec:main_proofs}. {  We can check the results in these theorems by conducting a simple numerical test: 1)  randomly generate a set of standard Gaussian vectors $\{\ba_k\}_{k=1}^m\subset\R^n$, 2) fix the measurement vectors and randomly {  generate} different $\bz$'s for each region, and 3) check whether the result in each theorem holds or not. For conciseness, we report in Table~\ref{table:main} the computational results in regions $\mathcal{R}_1$, $\mathcal{R}_{2c}$ and $\mathcal{R}_3$ (with three randomly generated $\bz$ in each region). } 

\begin{table}[t!]
\centering
\caption{  {  This table numerically checks the results in the theorems:} $\bx=[1,\cdots,1]^\top\in\R^{n}$ with $n=128$, $\{\ba_k\}_{k=1}^m\subset\R^n$ ($m=6n$) are independent standard  Gaussian vectors. The vectors $\bz$ in $\mathcal{R}_1$, $\mathcal{R}_{2c}$, and $\mathcal{R}_3$ are generated uniformly at random  (for $\mathcal{R}_{2c}$ we consider its intersection with the ball  $\{\bz:~\|\bz\|\leq2\|\bx\|\}$). The parameter $\delta$ which defines $\mathcal{R}_3$ is chosen to be $1/100$. The theoretical bounds refer to those given in Thms. \ref{thm:R1}, \ref{thm:R2c}, and \ref{thm:R3}.  }
\label{table:main}
\makegapedcells
\setcellgapes{3pt}
\begin{tabular}{c|ccc|ccc|ccc}
\hline
 &  \multicolumn{3}{c|}{$\lambda_{\min}\lb\nabla^2f(\bz)\rb$ in $\mathcal{R}_1$} &  \multicolumn{3}{c|}{$\bz^\top\nabla f(\bz)$ in $\mathcal{R}_{2c}$} &  \multicolumn{3}{c}{$\bz^\top\nabla f(\bz)$ in $\mathcal{R}_3$} \\
 \hline
{Numerical results} & 94.43 & 96.39 & 108.83 & 2.87e5 &2.61e5 &6.87e6  & -2184.41 &-2203.39 &-95.00\\
\hline
\multirow{ 2}{*}{Theoretical bounds}& 5.12 & 5.12 & 5.12 & 2632.14 & 3233.31 &55528.5 & -73.63 & -81.98 & -2.55\\
 &  \multicolumn{3}{c|}{{  (lower bound)}} &  \multicolumn{3}{c|}{{  (lower bound)}} &  \multicolumn{3}{c}{{  (upper bound)}}
\\
\hline
 \end{tabular}
\end{table}

From the five theorems, it is evident that critical points of $f(\bz)$ can only occur in $\MR_1$ and $\MR_{2a}$, since at critical points one has $\nabla f(\bz)=0$. Noticing that $\pm\bx\in\MR_1$, $f(\bz)\geq 0$ and $f(\pm\bx)=0$,  by Theorem~\ref{thm:R1},  we know that $\pm\bx$ are the local minimizers.  Theorem~\ref{thm:R2a} implies that at any critical point in $\MR_{2a}$, the Hession of $f(\bz)$ has a negative directional curvature as well as a positive directional curvature. Thus, critical points in $\MR_{2a}$ must be ridable saddle points \cite{SQW:FCM:18}. Putting it all together, we can establish Theorem~\ref{thm:main} and show that every local minimizer is a global minimizer. 
Additionally, though $f(\bz)$ is singular at $\bz=0$, Theorem~\ref{thm:R3} shows that local minimizers of $f(\bz)$ cannot exist around $0$. Moreover, it also implies that searching along the gradient descent direction at any point  in $\MR_{3}$ will move the point further away from the origin.
\section{Proofs for Section~\ref{sec:results}}\label{sec:main_proofs}
\subsection{Technical lemmas}
In order to prove the main theorems, we first list several technical lemmas that will be used repeatedly in this section, but defer the proofs to Section~\ref{sec:tech_proofs}.
Here and throughout this paper, if the expression of a random variable or a random matrix is long we will simply use $\repE$ to denote the associated expectation. 
\begin{lemma}\label{lem:tech_lemma1}
Let $h(u)$ be a continuous function defined on $[0,\infty)$ which obeys 
\begin{align*}
\begin{cases}
h(u) = 1 & \mbox{if }0\leq u\leq \beta,\\
h(u)\in[0,1] &\mbox{if }u\in{ (\beta,\gamma)},\\
h(u)=0&\mbox{if }u\geq \gamma
\end{cases}
\end{align*}
for two absolute numerical constants $\gamma>\beta\geq 1$. Assume $\ba_k\sim\N(0,\BI_n)$, $k=1,\cdots,m$, are independent. Then for any $\epsilon\in(0,1)$ and all nonzero vectors $\bu,~\bv\in\R^n$,
\begin{align*}
&\ln\frac{1}{m}\sum_{k=1}^m(\ba_k^\top\bu)^s(\ba_k^\top\bv)^th\lb\frac{|\ba_k^\top\bu|^2}{\|\bu\|^2}\rb h\lb\frac{|\ba_k^\top\bv|^2}{\|\bv\|^2}\rb\ba_k\ba_k^\top-\repE\rn\\
&\lesssim \lb\epsilon\cdot \max\{s,t\}\gamma^{\frac{t+s}{2}}+\gamma^{\frac{s+t}{2}} \epsilon^{-1}e^{-0.49\epsilon^{-2}}+\gamma^{\frac{s+t+1}{2}}e^{-0.49\beta}\rb\|\bu\|^s\|\bv\|^t
\end{align*}
holds with probability at least $1-e^{-\Omega(m\epsilon^2)}$ provided $m\gtrsim \epsilon^{-2}\log\epsilon^{-1}\cdot n$, where the exponents $s$ and $t$ are two nonnegative integers.
\end{lemma}
{\begin{lemma}\label{lem:tech_lemma12}
Under the setup of Lemma~\ref{lem:tech_lemma1},  
\begin{align*}
&\ln\mean{(\ba_k^\top\bu)^s(\ba_k^\top\bv)^t\lb h\lb\frac{|\ba_k^\top\bu|^2}{\|\bu\|^2}\rb h\lb\frac{|\ba_k^\top\bv|^2}{\|\bv\|^2}\rb-1\rb\ba_k\ba_k^\top}\rn\lesssim  ((8s)!!)^{1/8}((8t)!!)^{1/8}\|\bu\|^s\|\bv\|^t\cdot e^{-0.25\beta} 
\end{align*}
holds for all $\|\bu\|\neq 0$ and $\|\bv\|\neq 0$.
\end{lemma}}
\begin{lemma}\label{lem:tech_lemma11}
Under the setup of Lemma~\ref{lem:tech_lemma1},
\begin{align*}
&\ln\frac{1}{m}\sum_{k=1}^m(\ba_k^\top\bz)^s(\ba_k^\top\bx)^t h\lb\frac{|\ba_k^\top\bz|^2}{\|\bz\|^2}\rb\lsb h\lb\frac{m|\ba_k^\top\bx|^2}{\|\by\|_1}\rb-h\lb\frac{|\ba_k^\top\bx|^2}{\|\bx\|^2}\rb\rsb\ba_k\ba_k^\top\rn\\& \lesssim 2^{\frac{t}{2}}\gamma^{\frac{s+t}{2}}\lb\sqrt{\beta}e^{-0.245\beta}+\epsilon\rb\|\bz\|^s\|\bx\|^t
\end{align*}
holds uniformly for all $\|\bz\|\neq 0$ with probability at least $1-e^{-\Omega(m\epsilon^2)}$ provided $m\gtrsim\epsilon^{-2}\log\epsilon^{-1}\cdot n$.
\end{lemma}
\begin{lemma}\label{lem:tech_lemma2}
Let $h(u)$ and { $g(u)$} be two continuous functions defined on $[0,\infty)$ { satisfying} 
\begin{align*}
\begin{cases}
h(u) = 1 & \mbox{if }0\leq u\leq \beta,\\
h(u)\in[0,1] &\mbox{if }u\in(\beta,\gamma),\\
h(u)=0&\mbox{if }u\geq \gamma
\end{cases}
\quad\mbox{and}\quad
\begin{cases}
g(u) = 0 & \mbox{if }u\in[0,\beta]\cup[\gamma,\infty], \\
\lab g(u)\rab\leq 1 & \mbox{if } \beta<u<\gamma
\end{cases}
\end{align*}
for two absolute numerical constants $\gamma>\beta\geq 1$. Assume $\ba_k\sim\N(0,\BI_n)$, $k=1,\cdots,m$, are independent. Then for any $\epsilon\in(0,1)$ and all nonzero vectors $\bz\in\R^n$,
\begin{align*}
&\ln\frac{1}{m}\sum_{k=1}^m(\ba_k^\top\bz)^s(\ba_k^\top\bx)^t g\lb\frac{|\ba_k^\top\bz|^2}{\|\bz\|^2}\rb h\lb\frac{m|\ba_k^\top\bx|^2}{\|\by\|_1}\rb\ba_k\ba_k^\top\rn\lesssim  2^{\frac{t}{2}}\gamma^{\frac{s+t}{2}}\lb\sqrt{\beta}e^{-0.49\beta}+\epsilon\rb\|\bz\|^s\|\bx\|^t
\end{align*}
holds with probability at least $1-e^{-\Omega(m\epsilon^2)}$ provided $m\gtrsim \epsilon^{-2}\log\epsilon^{-1}\cdot n$, where the exponents $s$ and $t$ are two nonnegative integers.
\end{lemma}
\begin{lemma}\label{lem:tech_lemma3}
Under the setup of Lemma~\ref{lem:tech_lemma2},
\begin{align*}
&\ln\frac{1}{m}\sum_{k=1}^m(\ba_k^\top\bz)^s(\ba_k^\top\bx)^t g\lb\frac{|\ba_k^\top\bz|^2}{\|\bz\|^2}\rb h\lb\frac{m|\ba_k^\top\bx|^2}{\|\by\|_1}\rb\bz\ba_k^\top\rn\lesssim  2^{\frac{t}{2}}\
\gamma^{\frac{s+t+1}{2}}\lb e^{-0.245\beta}+\sqrt{\epsilon}\rb\|\bz\|^{s+1}\|\bx\|^t
\end{align*}
holds uniformly for all $\|\bz\|\neq 0$ with probability at least $1-e^{-\Omega(m\epsilon^2)}$ provided $m\gtrsim \epsilon^{-2}\log\epsilon^{-1}\cdot n$.
\end{lemma}
\begin{lemma}\label{lem:tech_lemma4}
Under the setup of Lemma~\ref{lem:tech_lemma2}, for $s\geq 2$
{\begin{align*}
&\lab\frac{1}{m}\sum_{k=1}^m(\ba_k^\top\bz)^s(\ba_k^\top\bx)^t g\lb\frac{|\ba_k^\top\bz|^2}{\|\bz\|^2}\rb h\lb\frac{m|\ba_k^\top\bx|^2}{\|\by\|_1}\rb \rab  \lesssim   2^{\frac{t}{2}}\gamma^{\frac{s+t}{2}}\lb\sqrt{\beta}e^{-0.49\beta}+\epsilon\rb
\|\bz\|^s\|\bx\|^t 
\end{align*}}
holds uniformly for all $\|\bz\|\neq 0$ with probability at least $1-e^{-\Omega(m\epsilon^2)}$ provided $m\gtrsim \epsilon^{-2}\log\epsilon^{-1}\cdot n$.
\end{lemma}
{\subsection{Proof of Theorem~\ref{thm:R1}}}
Due to symmetry, it suffices to consider the  region  ${ \|\bz-\bx\|}\leq \frac{1}{5}\|\bx\|$, from which we have \begin{align*}\frac{4}{5}\|\bx\|\leq\|\bz\|\leq\frac{6}{5}\|\bx\|.\numberthis\label{eq:thmR1_eq1}
\end{align*}
Though there are twelve terms in the expression for $\hessian f(\bz)$ (see \eqref{eq:hessian}), it is not difficult to see that the second term through the last term, with their sum denoted by $\TM_2$, can be bounded by Lemmas~\ref{lem:tech_lemma2} to \ref{lem:tech_lemma4} { (the details are deferred to Appendix \ref{sec:app:sub3})}, giving
\begin{align*}
\|\TM_2\| &\lesssim \gamma^{\frac{9}{2}}\lb e^{-0.245\beta}+\sqrt{\epsilon}\rb\max\lcb |h'|_{\infty},|h''|_{\infty}\rcb\max\lcb\|\bz\|^2,\|\bx\|^2,\frac{\|\bx\|^4}{\|\bz\|^2}\rcb\numberthis\label{eq:thmR1_eq2a0}\\
&\lesssim \gamma^{\frac{9}{2}}\lb e^{-0.245\beta}+\sqrt{\epsilon}\rb\max\lcb |h'|_{\infty},|h''|_{\infty}\rcb\|\bx\|^2,\numberthis\label{eq:thmR1_eq2}
\end{align*}
where we have used \eqref{eq:thmR1_eq1} in the second line. Define 
\begin{align*}
\TM_1 = \frac{1}{m}\sum_{k=1}^m\lb 6(\ba_k^\top\bz)^2-2(\ba_k^\top\bx)^2\rb h\lb\frac{|\ba_k^\top\bz|^2}{\|\bz\|^2}\rb h\lb\frac{m|\ba_k^\top\bx|^2}{\|\by\|_1}\rb\ba_k\ba_k^\top,
\end{align*}
which is the first term in the Hessian of $\hessian f(\bz)$. By setting $(s,t)$ to be $(2,0)$ and $(0,2)$ respectively in Lemma~\ref{lem:tech_lemma11}, we have 
\begin{align*}
&\ln\TM_1-\frac{1}{m}\sum_{k=1}^m\lb 6(\ba_k^\top\bz)^2-2(\ba_k^\top\bx)^2\rb h\lb\frac{|\ba_k^\top\bz|^2}{\|\bz\|^2}\rb h\lb\frac{|\ba_k^\top\bx|^2}{\|\bx\|^2}\rb\ba_k\ba_k^\top\rn\\
&\lesssim \gamma^{\frac{3}{2}}\lb e^{-0.245\beta}+\epsilon\rb\max\lcb\|\bz\|^2,\|\bx\|^2\rcb\\
&\lesssim \gamma^{\frac{3}{2}}\lb e^{-0.245\beta}+\epsilon\rb\|\bx\|^2.\numberthis\label{eq:thmR1_eq3}
\end{align*}
Moreover, letting $(s,t)$ to be $(2,0)$ and $(0,2)$ respectively in Lemma~\ref{lem:tech_lemma1}, we have 
\begin{align*}
&\lambda_{\min}\lb \frac{1}{m}\sum_{k=1}^m\lb 6(\ba_k^\top\bz)^2-2(\ba_k^\top\bx)^2\rb h\lb\frac{|\ba_k^\top\bz|^2}{\|\bz\|^2}\rb h\lb\frac{|\ba_k^\top\bx|^2}{\|\bx\|^2}\rb\ba_k\ba_k^\top\rb\\
&\geq \lambda_{\min}\lb\mean{\lb 6(\ba_k^\top\bz)^2-2(\ba_k^\top\bx)^2\rb h\lb\frac{|\ba_k^\top\bz|^2}{\|\bz\|^2}\rb h\lb\frac{|\ba_k^\top\bx|^2}{\|\bx\|^2}\rb\ba_k\ba_k^\top}\rb\\
&-C\gamma^{\frac{3}{2}}\lb\epsilon+ \epsilon^{-1}e^{-0.49\epsilon^{-2}}+e^{-0.49\beta}\rb\|\bx\|^2\\
&\geq \lambda_{\min}\lb\mean{\lb 6(\ba_k^\top\bz)^2-2(\ba_k^\top\bx)^2\rb\ba_k\ba_k^\top}\rb\\
&-\ln \mean{\lb 6(\ba_k^\top\bz)^2-2(\ba_k^\top\bx)^2\rb \lcb h\lb\frac{|\ba_k^\top\bz|^2}{\|\bz\|^2}\rb h\lb\frac{|\ba_k^\top\bx|^2}{\|\bx\|^2}\rb-1\rcb\ba_k\ba_k^\top}\rn\\
&-C\gamma^{\frac{3}{2}}\lb\epsilon+ \epsilon^{-1}e^{-0.49\epsilon^{-2}}+e^{-0.49\beta}\rb\|\bx\|^2,
\end{align*}
where $C$ is an absolute constant whose value may change from line to line. 

For any unit vector $\bq\in S^{n-1}$, we have 
\begin{align*}
&\bq^\top\mean{\lb 6(\ba_k^\top\bz)^2-2(\ba_k^\top\bx)^2\rb\ba_k\ba_k^\top}\bq\\
&=6\|\bq\|^2\|\bz\|^2+12(\bq^\top\bz)^2-2\|\bq\|^2\|\bx\|^2-4(\bq^\top\bx)^2\\
&\geq 6\|\bz\|^2-2\|\bx\|^2-4|\bq^\top(\bz+\bx)||\bq^\top(\bz-\bx)|\\
&\geq \frac{2}{25}\|\bx\|^2,
\end{align*}
{  where the second line follows from the standard result 
\begin{align*}
\mean{(a_k^\top\bu)^2(\ba_k^\top\bv)^2} = \|\bu\|^2\|\bv\|^2+2(\bu^\top\bv)^2,
\end{align*}
the third line follows from $12(\bq^\top\bz)^2\geq 4(\bq^\top\bz)^2$, and the last line follows from \eqref{eq:thmR1_eq1} which implies $\|\bz\|^2\geq \frac{16}{25}\|\bx\|^2$, $\|\bz+\bx\|\leq \frac{11}{5}\|\bx\|$, and $\|\bz-\bx\|\leq \frac{1}{5}\|\bx\|$.}
Moreover, the application of Lemma~\ref{lem:tech_lemma12} yields
{\begin{align*}
&\ln \mean{\lb 6(\ba_k^\top\bz)^2-2(\ba_k^\top\bx)^2\rb \lcb h\lb\frac{|\ba_k^\top\bz|^2}{\|\bz\|^2}\rb h\lb\frac{|\ba_k^\top\bx|^2}{\|\bx\|^2}\rb-1\rcb\ba_k\ba_k^\top}\rn\\
&\lesssim \lb\|\bz\|^2+\|\bx\|^2\rb{e^{-0.25\beta}}\\
&\lesssim e^{-0.25\beta}\|\bx\|^2,
\end{align*}}It follows that 
\begin{align*}
&\lambda_{\min}\lb \frac{1}{m}\sum_{k=1}^m\lb 6(\ba_k^\top\bz)^2-2(\ba_k^\top\bx)^2\rb h\lb\frac{|\ba_k^\top\bz|^2}{\|\bz\|^2}\rb h\lb\frac{|\ba_k^\top\bx|^2}{\|\bx\|^2}\rb\ba_k\ba_k^\top\rb\\
&\geq \lb \frac{2}{25}-Ce^{-0.25\beta}-C\gamma^{\frac{3}{2}}\lb\epsilon+ \epsilon^{-1}e^{-0.49\epsilon^{-2}}+e^{-0.49\beta}\rb\rb\|\bx\|^2.\numberthis\label{eq:thmR1_eq4}
\end{align*}
Noting that $\hessian f(\bz) = \TM_1+\TM_2$, combining \eqref{eq:thmR1_eq2}, \eqref{eq:thmR1_eq3}, and \eqref{eq:thmR1_eq4} together yields
\begin{align*}
\lambda_{\min}\lb\nabla^2 f(\bz)\rb&\geq \lb \frac{2}{25}-Ce^{-0.25\beta}-C\gamma^{\frac{3}{2}}\lb\epsilon+ \epsilon^{-1}e^{-0.49\epsilon^{-2}}+e^{-0.49\beta}\rb\rb\|\bx\|^2\\
&-C\gamma^{\frac{3}{2}}\lb e^{-0.245\beta}+\epsilon\rb\|\bx\|^2-C\gamma^{\frac{9}{2}}\lb e^{-0.245\beta}+\sqrt{\epsilon}\rb\max\lcb |h'|_{\infty},|h''|_{\infty}\rcb\|\bx\|^2\\
&\geq\frac{1}{25}\|\bx\|^2
\end{align*}
for sufficiently small $\epsilon$ and sufficiently large $\beta$ and $\gamma$ since $\max\lcb |h'|_{\infty},|h''|_{\infty}\rcb=O(1)$ in our construction.
\subsection{Proof of Theorem~\ref{thm:R2a}}
Due to symmetry we only need to consider the case $\bz^\top\bx\geq 0$ in $\MR_{2a}$. We will first show that with probability $1-e^{-\Omega(m)}$,
\begin{align*}
\bx^\top\nabla f(\bz)<-\frac{\delta}{100}\|\bx\|^4\numberthis\label{eq:R2a01}
\end{align*}
holds uniformly for all $\bz$ in the region $\MR_{2a}\cap \{\bz~|~\bz^\top\bx\geq \delta\|\bx\|^2\}$ provided $m\gtrsim n$, and hence excluding the possibility of any critical points in this region. 

Next, we will show that with probability $1-e^{-\Omega(m)}$, 
\begin{align*}
\bz^\top\nabla f(\bz) > \delta\|\bx\|^4\numberthis\label{eq:R2a02}
\end{align*}
holds uniformly for all $\bz$ in the region $\MR_{2a}\cap \{\bz~|~0\leq\bz^\top\bx< \delta\|\bx\|^2\}\cap \{\bz~|~\|\bz\|^2/\|\bx\|^2\geq\frac{1}{3}+\delta\}$ provided $m\gtrsim n$, and again excluding the possibility of any critical points in this region. 
Then the first part of Theorem~\ref{thm:R2a} follows immediately by combining the above two results together.

\paragraph{Proof of \eqref{eq:R2a01}} Notice that
\begin{align*}
\bx^\top\nabla f(\bz) & = \frac{1}{m}\sum_{k=1}^m2\lb (\ba_k^\top\bz)^2-(\ba_k^\top\bx)^2\rb(\ba_k^\top\bz)(\ba_k^\top\bx)h\lb\frac{|\ba_k^\top\bz|^2}{\|\bz\|^2}\rb h\lb\frac{m|\ba_k^\top\bx|^2}{\|\by\|_1}\rb\\
&+\frac{1}{\|\bz\|^2}\cdot\frac{1}{m}\sum_{k=1}^m\lb(\ba_k^\top\bz)^2-(\ba_k^\top\bx)^2\rb^2(\ba_k^\top\bz)(\ba_k^\top\bx)h'\lb\frac{|\ba_k^\top\bz|^2}{\|\bz\|^2}\rb h\lb\frac{m|\ba_k^\top\bx|^2}{\|\by\|_1}\rb\\
&-\frac{\bz^\top\bx}{\|\bz\|^4}\cdot\frac{1}{m}\sum_{k=1}^m\lb(\ba_k^\top\bz)^2-(\ba_k^\top\bx)^2\rb^2(\ba_k^\top\bz)^2h'\lb\frac{|\ba_k^\top\bz|^2}{\|\bz\|^2}\rb h\lb\frac{m|\ba_k^\top\bx|^2}{\|\by\|_1}\rb\\
&:=\TM_1+\TM_2+\TM_3.
\end{align*}

By Lemmas~\ref{lem:tech_lemma11}, \ref{lem:tech_lemma1},
and \ref{lem:tech_lemma12}, and noticing {$\|\bz\|^2\leq \|\bx\|^2$} in $\MR_{2a}$, we have 
\begin{align*}
\TM_1&\leq \frac{1}{m}\sum_{k=1}^m2\lb (\ba_k^\top\bz)^2-(\ba_k^\top\bx)^2\rb(\ba_k^\top\bz)(\ba_k^\top\bx)h\lb\frac{|\ba_k^\top\bz|^2}{\|\bz\|^2}\rb h\lb\frac{|\ba_k^\top\bx|^2}{\|\bz\|^2}\rb+C\gamma\lb\sqrt{\beta}e^{-0.245\beta}+\epsilon\rb\|\bx\|^4\\
&\leq \mean{2\lb (\ba_k^\top\bz)^2-(\ba_k^\top\bx)^2\rb(\ba_k^\top\bz)(\ba_k^\top\bx)h\lb\frac{|\ba_k^\top\bz|^2}{\|\bz\|^2}\rb h\lb\frac{|\ba_k^\top\bx|^2}{\|\bz\|^2}\rb}\\
&+C\lb\gamma\epsilon+\gamma\epsilon^{-1}e^{-0.49\epsilon^{-2}}+\gamma^{1.5}e^{-0.49\beta}\rb \|\bx\|^4+C\gamma\lb\sqrt{\beta}e^{-0.245\beta}+\epsilon\rb\|\bx\|^4\\
&\leq  \mean{2\lb (\ba_k^\top\bz)^2-(\ba_k^\top\bx)^2\rb(\ba_k^\top\bz)(\ba_k^\top\bx)}\\
&+Ce^{-0.25\beta}\|\bx\|^4+C\lb\gamma\epsilon+\gamma\epsilon^{-1}e^{-0.49\epsilon^{-2}}+\gamma^{1.5}e^{-0.49\beta}\rb \|\bx\|^4+C\gamma\lb\sqrt{\beta}e^{-0.245\beta}+\epsilon\rb\|\bx\|^4.
\end{align*}
On the other hand, by Lemma~\ref{lem:tech_lemma2} and noticing ${\frac{97}{300}\|\bx\|^2\leq(\frac{1}{3}-\delta)\|\bx\|^2 \leq \|\bz\|^2\leq \|\bx\|^2}$ in $\MR_{2a}$ since $\delta\leq\frac{1}{100}$, we have 
\begin{align*}
\TM_2+\TM_3\leq C|h'|_\infty\gamma^2\lb\sqrt{\beta}e^{-0.49\beta}+\epsilon\rb\|\bx\|^4.
\end{align*}
Thus, combining the above two inequalities together implies that for all
$\bz$ in  $\MR_{2a}\cap \{\bz~|~\bz^\top\bx\geq \delta\|\bx\|^2\}$ we have 
\begin{align*}
\bx^\top\nabla f(\bz)&\leq \mean{2\lb (\ba_k^\top\bz)^2-(\ba_k^\top\bx)^2\rb(\ba_k^\top\bz)(\ba_k^\top\bx)}+\frac{\delta}{20}\|\bx\|^4\\
&=6(\bz^\top\bx)(\|\bz\|^2-\|\bx\|^2)+\frac{\delta}{20}\|\bx\|^4\\
&\leq -\frac{\delta}{100}\|\bx\|^4
\end{align*}
where the first line can be achieved by choosing $\epsilon$ sufficiently small and $\gamma>\beta$ sufficiently large, and in the last line we have used the fact 
{$\|\bz\|^2\leq \frac{99}{100}\|\bx\|^2$ and $\bz^\top\bx\geq\delta\|\bx\|^2$} in
$\MR_{2a}\cap \{\bz~|~\bz^\top\bx\geq \delta\|\bx\|^2\}$.
\paragraph{Proof of \eqref{eq:R2a02}} First we have 
\begin{align*}
\bz^\top\nabla f(\bz) =\frac{1}{m}\sum_{k=1}^m2(|\ba_k^\top\bz|^2-|\ba_k^\top\bx|^2)(\ba_k^\top\bz)^2h\lb\frac{|\ba_k^\top\bz|^2}{\|\bz\|^2}\rb h\lb\frac{m|\ba_k^\top\bz|^2}{\|\by\|_1}\rb\numberthis\label{eq:ztfz}
\end{align*}
By applying Lemmas~\ref{lem:tech_lemma11}, \ref{lem:tech_lemma1} and \ref{lem:tech_lemma12} in order, we have 
\begin{align*}
\bz^\top\nabla f(\bz) &\geq 
\frac{1}{m}\sum_{k=1}^m2(|\ba_k^\top\bz|^2-|\ba_k^\top\bx|^2)(\ba_k^\top\bz)^2h\lb\frac{|\ba_k^\top\bz|^2}{\|\bz\|^2}\rb h\lb\frac{|\ba_k^\top\bz|^2}{\|\bx\|^2}\rb\\
&-C\gamma\lb\sqrt{\beta}e^{-0.245\beta}+\epsilon\rb\lb\|\bz\|^4+\|\bz\|^2\|\bx\|^2\rb\\
&\geq \mean{2(|\ba_k^\top\bz|^2-|\ba_k^\top\bx|^2)(\ba_k^\top\bz)^2h\lb\frac{|\ba_k^\top\bz|^2}{\|\bz\|^2}\rb h\lb\frac{|\ba_k^\top\bz|^2}{\|\bx\|^2}\rb}\\
&-C\gamma^{1.5}\lb 2\epsilon+\epsilon^{-1}e^{-0.49\epsilon^{-2}}+e^{-0.49\beta}\rb\lb\|\bz\|^4+\|\bz\|^2\|\bx\|^2\rb\\
&-C\gamma\lb\sqrt{\beta}e^{-0.245\beta}+\epsilon\rb\lb\|\bz\|^4+\|\bz\|^2\|\bx\|^2\rb\\
&\geq  \mean{2(|\ba_k^\top\bz|^2-|\ba_k^\top\bx|^2)(\ba_k^\top\bz)^2}-Ce^{-0.25\beta}\lb\|\bz\|^4+\|\bz\|^2\|\bx\|^2\rb\\
&-C\gamma^{1.5}\lb 2\epsilon+\epsilon^{-1}e^{-0.49\epsilon^{-2}}+e^{-0.49\beta}\rb\lb\|\bz\|^4+\|\bz\|^2\|\bx\|^2\rb\\
&-C\gamma\lb\sqrt{\beta}e^{-0.245\beta}+\epsilon\rb\lb\|\bz\|^4+\|\bz\|^2\|\bx\|^2\rb.\numberthis\label{eq:R2a03}
\end{align*}
Noticing that in $\MR_{2a}\cap \{\bz~|~0\leq\bz^\top\bx< \delta\|\bx\|^2\}\cap \{\bz~|~\|\bz\|^2/\|\bx\|^2\geq\frac{1}{3}+\delta\}$ we have $(\frac{1}{3}+\delta)\|\bx\|^2\leq\|\bz\|^2\leq\|\bx\|^2$, and consequently,
\begin{align*}
\bz^\top f(\bz) &\geq 6\|\bz\|^4-2\|\bz\|^2\|\bx\|^2-4(\bz^\top\bx)^2\\
&-\lb C\gamma\lb \sqrt{\beta}e^{-0.245\beta}+\epsilon\rb+C\gamma^{1.5}\lb 2\epsilon+C\epsilon^{-1}e^{-0.49\epsilon^{-2}}+e^{-0.49\beta}\rb+Ce^{-0.25\beta}\rb\|\bx\|^4\\
&\geq 6\|\bz\|^4-2\|\bz\|^2\|\bx\|^2-4(\bz^\top\bx)^2-\delta\|\bx\|^4\\
& \geq 6\lb\frac13+\delta\rb^2\|\bx\|^4-2\lb\frac13+\delta\rb\|\bx\|^4-4\delta^2\|\bx\|^4-\delta\|\bx\|^4\\
&=(\delta+2\delta^2)\|\bx\|^4>\delta\|\bx\|^4,
\end{align*}
where the second inequality can be achieved by choosing  $\epsilon$ to be  sufficiently small and $\gamma>\beta$ to be sufficiently large.\\

In the first part we have established that critical points in $\MR_{2a}$ must obey 
\begin{align*}
\frac{1}{3}-\delta<\frac{\|\bz\|^2}{\|\bx\|^2}<\frac{1}{3}+\delta\quad\mbox{and}\quad
|\bz^\top\bx|<\delta\|\bx\|^2.
\end{align*}
Thus, by \eqref{eq:thmR1_eq2}, we have 
\begin{align*}
\bx^\top\nabla^2 f(\bz)\bx &\leq \frac{1}{m}\sum_{k=1}^m\lb 6(\ba_k^\top\bz)^2-2(\ba_k^\top\bx)^2\rb(\ba_k^\top\bx)^2 h\lb\frac{|\ba_k^\top\bz|^2}{\|\bz\|^2}\rb h\lb\frac{m|\ba_k^\top\bx|^2}{\|\by\|_1}\rb\\
&+C\gamma^{\frac{9}{2}}\lb e^{-0.245\beta}+\sqrt{\epsilon}\rb\max\lcb |h'|_{\infty},|h''|_{\infty}\rcb\|\bx\|^4
\end{align*}
Applying Lemmas~\ref{lem:tech_lemma11}, \ref{lem:tech_lemma1} and \ref{lem:tech_lemma12} in order yields
\begin{align*}
\bx^\top\nabla^2f(\bz)\bx & \leq \mean{6(\ba_k^\top\bz)^2(\ba_k^\top\bx)^2-2(\ba_k^\top\bx)^4}\\
&+\lb C\gamma\lb \sqrt{\beta}e^{-0.245\beta}+\epsilon\rb+C\gamma^{1.5}\lb 2\epsilon+C\epsilon^{-1}e^{-0.49\epsilon^{-2}}+e^{-0.49\beta}\rb+Ce^{-0.25\beta}\rb\|\bx\|^4\\
&+C\gamma^{\frac{9}{2}}\lb e^{-0.245\beta}+\sqrt{\epsilon}\rb\max\lcb |h'|_{\infty},|h''|_{\infty}\rcb\|\bx\|^4\\
&\leq 6\|\bz\|^2\|\bx\|^2+12(\bz^\top\bx)^2-6\|\bx\|^4+\delta\|\bx\|^4\\
&\leq \lb6\lb\frac{1}{3}+\delta\rb+12\delta^2-6+\delta\rb\|\bx\|^4\\
&\leq -3\|\bx\|^4,
\end{align*}
where in the second inequality for fixed $\delta$ we choose $\epsilon$ to be sufficiently small and $\beta$ and $\gamma$ to be properly large.
Similarly, but considering a different direction, we have 
\begin{align*}
\bz^\top\nabla^2 f(\bz)\bz&\geq 18\|\bz\|^4-2\|\bz\|^2\|\bx\|^2-4(\bz^\top\bx)^2\\
&-\lb C\gamma\lb \sqrt{\beta}e^{-0.245\beta}+\epsilon\rb+C\gamma^{1.5}\lb 2\epsilon+C\epsilon^{-1}e^{-0.49\epsilon^{-2}}+e^{-0.49\beta}\rb+Ce^{-0.25\beta}\rb\|\bx\|^4\\
&-C\gamma^{\frac{9}{2}}\lb e^{-0.245\beta}+\sqrt{\epsilon}\rb\max\lcb |h'|_{\infty},|h''|_{\infty}\rcb\|\bx\|^4\\
&\geq \lb18\lb\frac{1}{3}-\delta\rb^2-2\lb\frac{1}{3}+\delta\rb-4\delta^2-\delta\rb\|\bx\|^4\\
&\geq \|\bx\|^4.
\end{align*}
\subsection{Proof of Theorem~\ref{thm:R2b}}
We only need to consider the case $\bz^\top\bx\geq 0$. Since in $\MR_{2b}$, one has 
\begin{align*}
\frac{99}{100}\leq \frac{\|\bz\|^2}{\|\bx\|^2}\leq\frac{101}{100}\quad \mbox{and} \quad\|\bz-\bx\|>\frac{1}{5}\|\bx\|.
\end{align*}
Thus, 
\begin{align*}
\bz^\top\bx\leq \frac{1}{2}\lb\|\bx\|^2+\|\bz\|^2-\frac{1}{25}\|\bx\|^2\rb\leq 0.985\|\bx\|^2.
\end{align*}
Noticing {$\|\bz\|^2\geq\frac{99}{100}\|\bx\|^2$} in $\MR_{2b}$, by choosing $\epsilon$ to be sufficiently small and $\beta$ and $\gamma$ to be properly large in \eqref{eq:R2a03}, we have 
\begin{align*}
\bz^\top\nabla f(\bz)&\geq \mean{2(|\ba_k^\top\bz|^2-|\ba_k^\top\bx|^2)(\ba_k^\top\bz)^2}-\delta\|\bx\|^4\\
&=6\|\bz\|^4-2\|\bz\|^2\|\bx\|^2-4(\bz^\top\bx)^2-\delta\|\bx\|^4\\
&=\|\bx\|^4\lb 6\lb\frac{\|\bz\|^2}{\|\bx\|^2}\rb^2-2\frac{\|\bz\|^2}{\|\bx\|^2}\rb-4(\bz^\top\bx)^2-\delta\|\bx\|^4\\
&\geq \lb6\cdot \frac{ 99^2}{100^2}-2\cdot \frac{99}{100}-4\cdot0.985^2-\delta\rb\|\bx\|^4\\
&\geq \frac{9}{1000}\|\bx\|^4
\end{align*}
provided {$\delta\leq \frac{1}{100}$}, where in the fourth line we have used the fact that the minimum of  $6\|\bz\|^4-2\|\bz\|^2\|\bx\|^2$ over $\frac{99}{100}\leq \frac{\|\bz\|^2}{\|\bx\|^2}$ is achieved at $\|\bz\|^2=\frac{99}{100}\|\bx\|^2$.
\subsection{Proof of Theorem~\ref{thm:R2c}}
Similarly to the proof for Theorem~\ref{thm:R2b}, we have 
\begin{align*}
\bz^\top\nabla f(\bz)&\geq 6\|\bz\|^4-2\|\bz\|^2\|\bx\|^2-4(\bz^\top\bx)^2-\delta\|\bx\|^4\\
&\geq 6\|\bz\|^2(\|\bz\|^2-\|\bx\|^2)-\delta\|\bx\|^4\\
&\geq  \frac{6}{101}\|\bz\|^4-\delta\|\bz\|^4\\
&\geq \frac{49}{1000}\|\bz\|^4,
\end{align*}
where in the third line we have used the fact {$\|\bz\|^2\geq\frac{101}{100}\|\bx\|^2$} in $\MR_{3c}$, and in the last line we have used the assumption {$\delta\leq\frac{1}{100}$}.
\subsection{Proof of Theorem~\ref{thm:R3}}
Recall that $\bz^\top\nabla f(\bz)$ is given in \eqref{eq:ztfz}.
Thus similar to \eqref{eq:R2a03} but applying by Lemmas~\ref{lem:tech_lemma11}, \ref{lem:tech_lemma1} and \ref{lem:tech_lemma12} in the reverse direction yields
\begin{align*}
\bz^\top\nabla f(\bz) 
&\leq \mean{2(|\ba_k^\top\bz|^4-|\ba_k^\top\bz|^2|\ba_k^\top\bx|^2)}+Ce^{-0.25\beta}\lb\|\bz\|^4+\|\bz\|^2\|\bx\|^2\rb\\
&+C\gamma^{1.5}\lb 2\epsilon+\epsilon^{-1}e^{-0.49\epsilon^{-2}}+e^{-0.49\beta}\rb\lb\|\bz\|^4+\|\bz\|^2\|\bx\|^2\rb\\
&+C\gamma\lb \sqrt{\beta}e^{-0.245\beta}+\epsilon\rb\lb\|\bz\|^4+\|\bz\|^2\|\bx\|^2\rb
\end{align*}
It follows that 
\begin{align*}
&\bz^\top\nabla f(\bz)\leq 6\|\bz\|^4-2\|\bz\|^2\|\bx\|^2-4(\bz^\top\bx)^2\\
&+\lb C\gamma\lb \sqrt{\beta}e^{-0.245\beta}+\epsilon\rb+C\gamma^{1.5}\lb 2\epsilon+C\epsilon^{-1}e^{-0.49\epsilon^{-2}}+e^{-0.49\beta}\rb+Ce^{-0.25\beta}\rb(\|\bz\|^4+\|\bz\|^2\|\bx\|^2)\\
&\leq 6\|\bz\|^4-2\|\bz\|^2\|\bx\|^2+\lb C\gamma\lb \sqrt{\beta}e^{-0.245\beta}+\epsilon\rb+C\gamma^{1.5}\lb 2\epsilon+C\epsilon^{-1}e^{-0.49\epsilon^{-2}}+e^{-0.49\beta}\rb+Ce^{-0.25\beta}\rb\|\bz\|^2\|\bx\|^2\\
&= \lcb 2\lb\frac{3\|\bz\|^2}{\|\bx\|^2}-1\rb+\lb C\gamma\lb \sqrt{\beta}e^{-0.245\beta}+\epsilon\rb+C\gamma^{1.5}\lb 2\epsilon+C\epsilon^{-1}e^{-0.49\epsilon^{-2}}+e^{-0.49\beta}\rb+Ce^{-0.25\beta}\rb\rcb\|\bz\|^2\|\bx\|^2\\
&\leq \lcb -6\delta+\lb C\gamma\lb \sqrt{\beta}e^{-0.245\beta}+\epsilon\rb+C\gamma^{1.5}\lb 2\epsilon+C\epsilon^{-1}e^{-0.49\epsilon^{-2}}+e^{-0.49\beta}\rb+Ce^{-0.25\beta}\rb\rcb\|\bz\|^2\|\bx\|^2\\
&\leq -5\delta\|\bz\|^2\|\bx\|^2,
\end{align*}
where in the second and the third inequalities  we have used the assumption { $\frac{\|\bz\|^2}{\|\bx\|^2}\leq \frac{1}{3}-\delta$}, and in 
the last inequality we choose $\epsilon$ to be  sufficiently small and $\gamma>\beta$ to be sufficiently large.

\section{Proofs of technical lemmas}\label{sec:tech_proofs}
\subsection{Proof of Lemma~\ref{lem:tech_lemma1}}
Due to the homogeneity, it suffices to establish the inequality for all $\bu\in\S^{n-1}$ and $\bv\in\S^{n-1}$. We will first consider a fixed pair of $\bu$ and $\bv$ and then use the covering argument. For fixed $\bu$ and $\bv$ of unit norm, it suffices to establish a uniform bound for 
\begin{align*}
\lab \frac{1}{m}\sum_{k=1}^m(\ba_k^\top\bu)^s(\ba_k^\top\bv)^th\lb|\ba_k^\top\bu|^2\rb h\lb|\ba_k^\top\bv|^2\rb(\ba_k^\top\bw)^2-\repE\rab\numberthis\label{eq:tech1_eq1}
\end{align*}
over all $\bw\in\N_{1/4}$, where $\N_{1/4}$ is a $1/4$-net of $\S^{n-1}$. { This is because for any symmetric matrix $\BA$ one has 
\begin{align*}
\|\BA\|\leq 2\sup_{\bz\in\N_{1/4}}|\langle \BA\bz,\bz\rangle|,
\end{align*} 
see Lemma~5.4 in \cite{Ver2011rand}}. Noticing that 
\begin{align*}
&\lab(\ba_k^\top\bu)^s(\ba_k^\top\bv)^th\lb|\ba_k^\top\bu|^2\rb h\lb|\ba_k^\top\bv|^2\rb(\ba_k^\top\bw)^2\rab\\
&\leq |\ba_k^\top\bu|^s|\ba_k^\top\bv|^t\dsone{|\ba_k^\top\bu|^2\leq\gamma} \dsone{|\ba_k^\top\bv|^2\leq\gamma}(\ba_k^\top\bw)^2\\
&\leq \gamma^{\frac{s+t}{2}}(\ba_k^\top\bw)^2
\end{align*}
and $(\ba_k^\top\bw)^2$ is a standard Chi-square, we can see that \begin{align*}(\ba_k^\top\bu)^s(\ba_k^\top\bv)^th\lb|\ba_k^\top\bu|^2\rb h\lb|\ba_k^\top\bv|^2\rb(\ba_k^\top\bw)^2
\end{align*}
is sub-exponential with the sub-exponential norm $\|\cdot\|_{\psi_1}$ bounded by an absolute  constant times $\gamma^{\frac{s+t}{2}}$. It follows that \cite{Ver2011rand}
\begin{align*}
\ln(\ba_k^\top\bu)^s(\ba_k^\top\bv)^th\lb|\ba_k^\top\bu|^2\rb h\lb|\ba_k^\top\bv|^2\rb(\ba_k^\top\bw)^2
-\repE\rn_{\psi_1}\lesssim \gamma^{\frac{s+t}{2}}.
\end{align*}
Thus the application of the Bernstein's inequality implies that 
\begin{align*}
\lab \frac{1}{m}\sum_{k=1}^m(\ba_k^\top\bu)^s(\ba_k^\top\bv)^th\lb|\ba_k^\top\bu|^2\rb h\lb|\ba_k^\top\bv|^2\rb(\ba_k^\top\bw)^2-\repE\rab\lesssim \gamma^{\frac{s+t}{2}}\epsilon\numberthis\label{eq:tech1_eq2}
\end{align*}
with probability at least $1-2e^{-\Omega(m\epsilon^2)}$ for $\epsilon\in(0,1)$. 
{  By \cite[Lemma 5.2]{Ver2011rand} we know that $|\N_{1/4}|\leq 9^n$. Thus the failure probability over all $\bw\in\N_{1/4}$ can be bounded by $9^n\cdot 2e^{-\Omega(m\epsilon^2)}=2 e^{-\Omega(m\epsilon^2)+n\log9}$ which is less than $2e^{-\Omega(m\epsilon^2)}$ (with a different constant hidden in $\Omega(m\epsilon^2)$) provided $m\gtrsim \epsilon^{-2}\cdot n$.}
Therefore, 
\begin{align*}
&\ln\frac{1}{m}\sum_{k=1}^m(\ba_k^\top\bu)^s(\ba_k^\top\bv)^th\lb|\ba_k^\top\bu|^2\rb h\lb|\ba_k^\top\bv|^2\rb\ba_k\ba_k^\top-\repE\rn\lesssim \gamma^{\frac{s+t}{2}}\epsilon\numberthis\label{eq:tech1_eq3}
\end{align*}
for fixed $\bu\in\S^{n-1}$ and $\bv\in\S^{n-1}$ with probability at least 
$1-2e^{-\Omega(m\epsilon^2)}$ provided $m\gtrsim \epsilon^{-2}\cdot n$.

To establish a bound over all $\bu\in\S^{n-1}$ and $\bv\in\S^{n-1}$, we will use the covering argument again. Let $\N_{\epsilon^2}$ be a $\epsilon^2$-net of $\S^{n-1}$ with cardinality $|\N_{\epsilon^2}|\leq (3/\epsilon^2)^n$. Then it is evident that \eqref{eq:tech1_eq3} holds for all $\bu_0\in\N_{\epsilon^2}$ and $\bv_0\in\N_{\epsilon^2}$ with probability at least 
$1-2e^{-\Omega(m\epsilon^2)}$ provided $m\gtrsim \epsilon^{-2}\log\epsilon^{-1}\cdot n$. For any $\bu\in\S^{n-1}$ and $\bv\in\S^{n-1}$, there exists a pair of $\bu_0,\bv_0\in\N_{\epsilon^2}$ such that $\|\bu-\bu_0\|\leq\epsilon^2$ and $\|\bv-\bv_0\|\leq\epsilon^2$. It follows that 
\begin{align*}
&\ln\frac{1}{m}\sum_{k=1}^m\lcb(\ba_k^\top\bu)^s(\ba_k^\top\bv)^th\lb|\ba_k^\top\bu|^2\rb h\lb|\ba_k^\top\bv|^2\rb\ba_k\ba_k^\top-(\ba_k^\top\bu_0)^s(\ba_k^\top\bv_0)^th\lb|\ba_k^\top\bu_0|^2\rb h\lb|\ba_k^\top\bv_0|^2\rb\ba_k\ba_k^\top\rcb\rn\\
&\leq \ln\frac{1}{m}\sum_{k=1}^m\lcb (\ba_k^\top\bu)^sh\lb|\ba_k^\top\bu|^2\rb-(\ba_k^\top\bu_0)^sh\lb|\ba_k^\top\bu_0|^2\rb\rcb(\ba_k^\top\bv)^th\lb|\ba_k^\top\bv|^2\rb\ba_k\ba_k^\top\rn\\
&+\ln\frac{1}{m}\sum_{k=1}^m(\ba_k^\top\bu_0)^sh\lb|\ba_k^\top\bu_0|^2\rb\lcb (\ba_k^\top\bv)^th\lb|\ba_k^\top\bv|^2\rb-(\ba_k^\top\bv_0)^th\lb|\ba_k^\top\bv_0|^2\rb\rcb\ba_k\ba_k^\top\rn.\numberthis\label{eq:tech1_eq4}
\end{align*}
Next we will focus on the first term of \eqref{eq:tech1_eq4} and the second term can be similarly bounded. 
We can split the first term into five terms based on the decomposition of $[0,\infty)\times [0,\infty)$ and then provide an upper bound for each term. 
\paragraph{Region $[0,\beta]\times[0,\beta]$}
\begin{align*}
&\ln\frac{1}{m}\sum_{k=1}^m\lcb (\ba_k^\top\bu)^sh\lb|\ba_k^\top\bu|^2\rb-(\ba_k^\top\bu_0)^sh\lb|\ba_k^\top\bu_0|^2\rb\rcb\dsone{{ |\ba_k^\top\bu|^2\leq\beta,|\ba_k^\top\bu_0|^2\leq\beta}}(\ba_k^\top\bv)^th\lb|\ba_k^\top\bv|^2\rb\ba_k\ba_k^\top\rn\\
&\leq \ln\frac{1}{m}\sum_{k=1}^m\lcb (\ba_k^\top\bu)^s-(\ba_k^\top\bu_0)^s\rcb\dsone{|\ba_k^\top\bu|\leq{ \sqrt{\beta}},|\ba_k^\top\bu_0|\leq{ \sqrt{\beta}},|\ba_k^\top\bu-\ba_k^\top\bu_0|\leq\epsilon}(\ba_k^\top\bv)^th\lb|\ba_k^\top\bv|^2\rb\ba_k\ba_k^\top\rn\\
&+\ln\frac{1}{m}\sum_{k=1}^m\lcb (\ba_k^\top\bu)^s-(\ba_k^\top\bu_0)^s\rcb\dsone{|\ba_k^\top\bu|\leq{ \sqrt{\beta}},|\ba_k^\top\bu_0|\leq{ \sqrt{\beta}},|\ba_k^\top\bu-\ba_k^\top\bu_0|>\epsilon}(\ba_k^\top\bv)^th\lb|\ba_k^\top\bv|^2\rb\ba_k\ba_k^\top\rn\\
&=\ln\frac{1}{m}\sum_{k=1}^m\lcb (\ba_k^\top\bu-\ba_k^\top\bu_0)\lb(\ba_k^\top\bu)^{s-1}+(\ba_k^\top\bu)^{s-2}(\ba_k^\top\bu_0)+\cdots+(\ba_k^\top\bu_0)^{s-1}\rb\rcb\right.\\&\left.{\color{white}\sum_{k=1}^m}\dsone{|\ba_k^\top\bu|\leq{ \sqrt{\beta}},|\ba_k^\top\bu_0|\leq{ \sqrt{\beta}},|\ba_k^\top\bu-\ba_k^\top\bu_0|\leq\epsilon}(\ba_k^\top\bv)^th\lb|\ba_k^\top\bv|^2\rb\ba_k\ba_k^\top\rn\\
&+\ln\frac{1}{m}\sum_{k=1}^m\lcb (\ba_k^\top\bu)^s-(\ba_k^\top\bu_0)^s\rcb\dsone{|\ba_k^\top\bu|\leq{ \sqrt{\beta}},|\ba_k^\top\bu_0|\leq{ \sqrt{\beta}},|\ba_k^\top\bu-\ba_k^\top\bu_0|>\epsilon}(\ba_k^\top\bv)^th\lb|\ba_k^\top\bv|^2\rb\ba_k\ba_k^\top\rn\\
&\leq \epsilon\cdot s\beta^{\frac{s-1}{2}}\gamma^{\frac{t}{2}}\ln\frac{1}{m}\sum_{k=1}^m\ba_k\ba_k^\top\rn+2\beta^{\frac{s}{2}}\gamma^{\frac{t}{2}}\ln\frac{1}{m}\sum_{k=1}^m\ba_k\ba_k^\top\dsone{|\ba_k^\top\bu-\ba_k^\top\bu_0|>\epsilon}\rn\\
&\leq \epsilon\cdot s\beta^{\frac{s-1}{2}}\gamma^{\frac{t}{2}}\ln\frac{1}{m}\sum_{k=1}^m\ba_k\ba_k^\top\rn+2\beta^{\frac{s}{2}}\gamma^{\frac{t}{2}}\ln\frac{1}{m}\sum_{k=1}^m\ba_k\ba_k^\top\dsone{|\ba_k^\top\bu-\ba_k^\top\bu_0|>\epsilon^{-1}\|\bu-\bu_0\|}\rn,
\end{align*}
{ where in the second inequality we have used the fact that because $h\lb|\ba_k^\top\bv|^2\rb=0$ when $|\ba_k^\top\bv|^2\geq\gamma$ there holds
\begin{align*}
(\ba_k^\top\bv)^th\lb|\ba_k^\top\bv|^2\rb\leq \gamma^{t/2},
\end{align*}}
and the last equality follows from the assumption $\|\bu-\bu_0\|\leq\epsilon^2$. {
Note that in the above calculation, it requires $s\geq 1$. However, when $s=0$, we have 
\begin{align*}
\frac{1}{m}\sum_{k=1}^m\lcb (\ba_k^\top\bu)^sh\lb|\ba_k^\top\bu|^2\rb-(\ba_k^\top\bu_0)^sh\lb|\ba_k^\top\bu_0|^2\rb\rcb\dsone{|\ba_k^\top\bu|\leq{ \sqrt{\beta}},|\ba_k^\top\bu_0|\leq{ \sqrt{\beta}}}(\ba_k^\top\bv)^th\lb|\ba_k^\top\bv|^2\rb\ba_k\ba_k^\top=0,
\end{align*}
and hence the upper bound still holds.
}
\paragraph{Region $[0,\beta]\times(\beta,\gamma]$ or $(\beta,\gamma]\times [0,\beta]$}
\begin{align*}
&\ln\frac{1}{m}\sum_{k=1}^m\lcb (\ba_k^\top\bu)^sh\lb|\ba_k^\top\bu|^2\rb-(\ba_k^\top\bu_0)^sh\lb|\ba_k^\top\bu_0|^2\rb\rcb\dsone{|\ba_k^\top\bu|^2\leq\beta,\beta<|\ba_k^\top\bu_0|^2\leq\gamma}(\ba_k^\top\bv)^th\lb|\ba_k^\top\bv|^2\rb\ba_k\ba_k^\top\rn\\
&\leq \lb\beta^{\frac{s}{2}}+\gamma^{\frac{s}{2}}\rb\gamma^{\frac{t}{2}}
\ln\frac{1}{m}\sum_{k=1}^m\ba_k\ba_k^\top\dsone{|\ba_k^\top\bu_0|>\sqrt{\beta}\|\bu_0\|}\rn
\end{align*}
and
\begin{align*}
&\ln\frac{1}{m}\sum_{k=1}^m\lcb (\ba_k^\top\bu)^sh\lb|\ba_k^\top\bu|^2\rb-(\ba_k^\top\bu_0)^sh\lb|\ba_k^\top\bu_0|^2\rb\rcb\dsone{\beta<|\ba_k^\top\bu|^2\leq\gamma,|\ba_k^\top\bu_0|^2\leq\beta}(\ba_k^\top\bv)^th\lb|\ba_k^\top\bv|^2\rb\ba_k\ba_k^\top\rn\\
&\leq \lb\beta^{\frac{s}{2}}+\gamma^{\frac{s}{2}}\rb\gamma^{\frac{t}{2}}
\ln\frac{1}{m}\sum_{k=1}^m\ba_k\ba_k^\top\dsone{|\ba_k^\top\bu|>\sqrt{\beta}\|\bu\|}\rn.
\end{align*}
\paragraph{Region $[0,\beta]\times(\gamma,\infty)$ or $(\gamma,\infty)\times [0,\beta]$}
\begin{align*}
&\ln\frac{1}{m}\sum_{k=1}^m (\ba_k^\top\bu)^sh\lb|\ba_k^\top\bu|^2\rb\dsone{|\ba_k^\top\bu|^2\leq\beta,|\ba_k^\top\bu_0|^2>\gamma}(\ba_k^\top\bv)^th\lb|\ba_k^\top\bv|^2\rb\ba_k\ba_k^\top\rn\\
&\leq\beta^{\frac{s}{2}}\gamma^{\frac{t}{2}}\ln\frac{1}{m}\sum_{k=1}^m\ba_k\ba_k^\top\dsone{|\ba_k^\top\bu_0|>\sqrt{\gamma}\|\bu_0\|}\rn
\end{align*}
and
\begin{align*}
&\ln\frac{1}{m}\sum_{k=1}^m (\ba_k^\top\bu_0)^sh\lb|\ba_k^\top\bu_0|^2\rb\dsone{|\ba_k^\top\bu|^2>\gamma,|\ba_k^\top\bu_0|^2\leq\beta}(\ba_k^\top\bv)^th\lb|\ba_k^\top\bv|^2\rb\ba_k\ba_k^\top\rn\\
&\leq\beta^{\frac{s}{2}}\gamma^{\frac{t}{2}}\ln\frac{1}{m}\sum_{k=1}^m\ba_k\ba_k^\top\dsone{|\ba_k^\top\bu|>\sqrt{\gamma}\|\bu\|}\rn.
\end{align*}
\paragraph{Region $(\beta,\gamma]\times(\beta,\gamma]$}
\begin{align*}
&\ln\frac{1}{m}\sum_{k=1}^m\lcb (\ba_k^\top\bu)^sh\lb|\ba_k^\top\bu|^2\rb-(\ba_k^\top\bu_0)^sh\lb|\ba_k^\top\bu_0|^2\rb\rcb\dsone{\beta<|\ba_k^\top\bu|^2\leq\gamma,\beta<|\ba_k^\top\bu_0|^2\leq\gamma}(\ba_k^\top\bv)^th\lb|\ba_k^\top\bv|^2\rb\ba_k\ba_k^\top\rn\\
&\leq 2\gamma^{\frac{s+t}{2}}\ln\frac{1}{m}\sum_{k=1}^m\ba_k\ba_k^\top\dsone{|\ba_k^\top\bu|>\sqrt{\beta}\|\bu\|}\rn.
\end{align*}
\paragraph{Region $(\beta,\gamma]\times(\gamma,\infty)$ or $(\gamma,\infty)\times(\beta,\gamma]$}
\begin{align*}
&\ln\frac{1}{m}\sum_{k=1}^m (\ba_k^\top\bu)^sh\lb|\ba_k^\top\bu|^2\rb\dsone{\beta<|\ba_k^\top\bu|^2\leq\gamma,|\ba_k^\top\bu_0|^2>\gamma}(\ba_k^\top\bv)^th\lb|\ba_k^\top\bv|^2\rb\ba_k\ba_k^\top\rn\\
&\leq\gamma^{\frac{s+t}{2}}\ln\frac{1}{m}\sum_{k=1}^m\ba_k\ba_k^\top\dsone{|\ba_k^\top\bu_0|>\sqrt{\gamma}\|\bu_0\|}\rn
\end{align*}
and
\begin{align*}
&\ln\frac{1}{m}\sum_{k=1}^m (\ba_k^\top\bu_0)^sh\lb|\ba_k^\top\bu_0|^2\rb\dsone{|\ba_k^\top\bu|^2>\gamma,\beta<|\ba_k^\top\bu_0|^2\leq\gamma}(\ba_k^\top\bv)^th\lb|\ba_k^\top\bv|^2\rb\ba_k\ba_k^\top\rn\\
&\leq\gamma^{\frac{s+t}{2}}\ln\frac{1}{m}\sum_{k=1}^m\ba_k\ba_k^\top\dsone{|\ba_k^\top\bu|>\sqrt{\gamma}\|\bu\|}\rn.
\end{align*}\\

Combining the bounds from (a) to (e) together and noting that the second term in \eqref{eq:tech1_eq4} can be bounded similarly to the first one yields that 
{ \begin{align*}
&\ln\frac{1}{m}\sum_{k=1}^m\lcb(\ba_k^\top\bu)^s(\ba_k^\top\bv)^th\lb|\ba_k^\top\bu|^2\rb h\lb|\ba_k^\top\bv|^2\rb\ba_k\ba_k^\top-(\ba_k^\top\bu_0)^s(\ba_k^\top\bv_0)^th\lb|\ba_k^\top\bu_0|^2\rb h\lb|\ba_k^\top\bv_0|^2\rb\ba_k\ba_k^\top\rcb\rn\\
&\lesssim\underbrace{\epsilon\cdot s\beta^{\frac{s-1}{2}}\gamma^{\frac{t}{2}}+\beta^{\frac{s}{2}}\gamma^{\frac{t}{2}}\lb \epsilon^{-1}e^{-0.49\epsilon^{-2}}+\epsilon\rb}_{\mbox{bound for (a)}}+\underbrace{\lb\beta^{\frac{s}{2}}+\gamma^{\frac{s}{2}}\rb\gamma^{\frac{t}{2}}\lb\sqrt{\beta}e^{-0.49\beta}+\epsilon\rb}_{\mbox{bound for (b)}}+\underbrace{
\beta^{\frac{s}{2}}\gamma^{\frac{t}{2}}\lb\sqrt{\gamma}e^{-0.49\gamma}+\epsilon\rb}_{\mbox{bound for (c)}}\\
&\quad+\underbrace{\gamma^{\frac{s+t}{2}}\lb \sqrt{\beta}e^{-0.49\beta}+\epsilon\rb}_{\mbox{bound for (d)}}+\underbrace{\gamma^{\frac{s+t}{2}}\lb \sqrt{\gamma}e^{-0.49\gamma}+\epsilon\rb}_{\mbox{bound for (e)}}\\
&\lesssim\epsilon\cdot \max\{s,t\}\gamma^{\frac{t+s}{2}}+\gamma^{\frac{s+t}{2}} \epsilon^{-1}e^{-0.49\epsilon^{-2}}+\gamma^{\frac{s+t+1}{2}}e^{-0.49\beta},\numberthis\label{eq:tech1_eq5}
\end{align*}
where each term in the first inequality respectively corresponds  the bound for (a) to (e) after applying Lemmas~\ref{lem:aux_lemma1} and \ref{lem:aux_lemma2}, and in the second inequality we have used the fact $1<\beta < \gamma$.}

{
By the same splitting scheme, we can similarly show that 
\begin{align*}
&\ln\mean{(\ba_k^\top\bu)^s(\ba_k^\top\bv)^th\lb|\ba_k^\top\bu|^2\rb h\lb|\ba_k^\top\bv|^2\rb\ba_k\ba_k^\top}-\mean{(\ba_k^\top\bu_0)^s(\ba_k^\top\bv_0)^th\lb|\ba_k^\top\bu_0|^2\rb h\lb|\ba_k^\top\bv_0|^2\rb\ba_k\ba_k^\top}\rn\\
&\lesssim\epsilon\cdot \max\{s,t\}\gamma^{\frac{t+s}{2}}+\gamma^{\frac{s+t}{2}} \epsilon^{-1}e^{-0.5\epsilon^{-2}}+\gamma^{\frac{s+t+1}{2}}e^{-0.5\beta}.\numberthis\label{eq:tech1_eq6}
\end{align*}
}Then the proof is complete after combining \eqref{eq:tech1_eq3}, \eqref{eq:tech1_eq5} and \eqref{eq:tech1_eq6} together and using the triangular inequality.
\subsection{Proof of Lemma~\ref{lem:tech_lemma12}}
A direct calculation yields that 
\begin{align*}
&\ln\mean{(\ba_k^\top\bu)^s(\ba_k^\top\bv)^th\lb\frac{|\ba_k^\top\bu|^2}{\|\bu\|^2}\rb h\lb\frac{|\ba_k^\top\bv|^2}{\|\bv\|^2}\rb\ba_k\ba_k^\top}-\mean{(\ba_k^\top\bu)^s(\ba_k^\top\bv)^t\ba_k\ba_k^\top}\rn\\
&=\max_{\|\bq\|=1}\lab \mean{(\ba_k^\top\bu)^s(\ba_k^\top\bv)^t(\ba_k^\top\bq)^2\lb h\lb\frac{|\ba_k^\top\bu|^2}{\|\bu\|^2}\rb h\lb\frac{|\ba_k^\top\bv|^2}{\|\bv\|^2}\rb-1\rb}\rab\\
&\leq \max_{\|\bq\|=1}\mean{|\ba_k^\top\bu|^s|\ba_k^\top\bv|^t|\ba_k^\top\bq|^2\lb 1-h\lb\frac{|\ba_k^\top\bu|^2}{\|\bu\|^2}\rb h\lb\frac{|\ba_k^\top\bv|^2}{\|\bv\|^2}\rb\rb}\\
&\leq \max_{\|\bq\|=1}\mean{|\ba_k^\top\bu|^s|\ba_k^\top\bv|^t|\ba_k^\top\bq|^2\lb\dsone{|\ba_k^\top\bu|>\sqrt{\beta}\|\bu\|}+\dsone{|\ba_k^\top\bv|>\sqrt{\beta}\|\bv\|}\rb}\\
&\leq \max_{\|\bq\|=1}\lb\mean{\lb|\ba_k^\top\bu|^{2s}|\ba_k^\top\bv|^{2t}|\ba_k^\top\bq|^4\rb}\rb^{1/2}\lb\sqrt{\mean{\dsone{|\ba_k^\top\bu|>\sqrt{\beta}\|\bu\|}}}+\sqrt{\mean{\dsone{|\ba_k^\top\bv|>\sqrt{\beta}\|\bv\|}}}\rb\\
&{ {\lesssim \left(\max_{\|\bq\|=1}\mathbb{E}|\ba_k^\top\bq|^8\right)^{\frac{1}{4}}\left( \mathbb{E}|\ba_k^\top\bu|^{4s}|\ba_k^\top\bv|^{4t} \right)^{\frac{1}{4}} \sqrt{\sqrt{\frac{2}{\pi\beta}}e^{-\frac{\beta}{2}}}}}\\
&{ {\lesssim 
\left(\mathbb{E}|\ba_k^\top\bu|^{8s}\right)^{\frac{1}{8}} \left(\mathbb{E}|\ba_k^\top\bv|^{8t}\right)^{\frac{1}{8}}e^{-0.25\beta}}}\\
&\lesssim ((8s)!!)^{1/8}((8t)!!)^{1/8}\|\bu\|^s\|\bv\|^t\cdot e^{-0.25\beta},
\end{align*}
{ {where the fourth inequality follows from H\"older's inequality and the fact $$\mean{\dsone{\{|\ba_k^\top\bu|>\sqrt{\beta}\|\bu\|\}}}=2\int_{\sqrt{\beta}}^{\infty} \frac{1}{\sqrt{2\pi}} e^{-\frac{t^2}{2}}dt\leq \sqrt{\frac{2}{\pi}}\int_{\sqrt{\beta}}^{\infty}\frac{t}{\sqrt{\beta}}e^{-\frac{t^2}{2}}dt\leq \sqrt{\frac{2}{\pi\beta}}e^{-\frac{\beta}{2}}$$  as well as $\beta>1$, and the fifth and sixth inequalities hold as the $2k$-th moment of a standard Gaussian variable is $(2k)!!$.
}}
\subsection{Proof of Lemma~\ref{lem:tech_lemma11}}
{  Noting that 
\begin{align*}
\frac{1}{m}\|\by\|_1 = \frac{1}{m}\sum_{k=1}^m(\ba_k^\top\bx)^2=\lab\bx^\top \lb\frac{1}{m}\sum_{k=1}^m\ba_k\ba_k^\top\rb\bx\rab,
\end{align*}
it follows from Lemma~\ref{lem:aux_lemma1} that $\frac{1}{2}\|\bx\|^2\leq\frac{1}{m}\|\by\|_1\leq2\|\bx\|^2$ holds with probability $1-e^{\Omega(m)}$ provided $m\gtrsim n$.} Thus, on the same event, we have 
\begin{align*}
&\ln\frac{1}{m}\sum_{k=1}^m(\ba_k^\top\bz)^s(\ba_k^\top\bx)^t h\lb\frac{|\ba_k^\top\bz|^2}{\|\bz\|^2}\rb\lsb h\lb\frac{m|\ba_k^\top\bx|^2}{\|\by\|_1}\rb-h\lb\frac{|\ba_k^\top\bx|^2}{\|\bx\|^2}\rb\rsb\ba_k\ba_k^\top\rn\\
&\leq \ln\frac{1}{m}\sum_{k=1}^m|\ba_k^\top\bz|^s|\ba_k^\top\bx|^t h\lb\frac{|\ba_k^\top\bz|^2}{\|\bz\|^2}\rb\left|h\lb\frac{m|\ba_k^\top\bx|^2}{\|\by\|_1}\rb-h\lb\frac{|\ba_k^\top\bx|^2}{\|\bx\|^2}\rb\right|\ba_k\ba_k^\top\rn  \\
&\leq \ln\frac{1}{m}\sum_{k=1}^m|\ba_k^\top\bz|^s|\ba_k^\top\bx|^t h\lb\frac{|\ba_k^\top\bz|^2}{\|\bz\|^2}\rb\cdot\dsone{|\ba_k^\top\bz|^2<\gamma\|\bz\|^2}\cdot\dsone{\frac{\beta}{2}\|\bx\|^2\leq |\ba_k^\top\bx|^2\leq2\gamma\|\bx\|^2 }\ba_k\ba_k^\top\rn\\
&\leq \|\bz\|^s\|\bx\|^t\cdot2^{\frac{t}{2}}\gamma^{\frac{s+t}{2}}\cdot \ln\frac{1}{m}\sum_{k=1}^m \mathbf{1}_{\{|\ba_k^\top\bx|\geq\sqrt{\frac{\beta}{2}}\|\bx\| \}}\ba_k\ba_k^\top\rn\\
&\lesssim \|\bz\|^s\|\bx\|^t\cdot2^{\frac{t}{2}}\gamma^{\frac{s+t}{2}}\lb\sqrt{\beta}e^{-0.245\beta}+\epsilon\rb,
\end{align*}
where { in the third inequality we have used the facts 
\begin{align*}
|\ba_k^\top\bz|^sh\lb\frac{|\ba_k^\top\bz|^2}{\|\bz\|^2}\rb\cdot\dsone{|\ba_k^\top\bz|^2<\gamma\|\bz\|^2}\leq\gamma^{\frac{s}{2}}\|\bz\|^s
\end{align*}
and 
\begin{align*}
|\ba_k^\top\bx|^t\dsone{\frac{\beta}{2}\|\bx\|^2\leq |\ba_k^\top\bx|^2\leq2\gamma\|\bx\|^2 }\leq 2^{\frac{t}{2}}\gamma^{\frac{t}{2}} \|\bx\|^t\mathbf{1}_{\{|\ba_k^\top\bx|\geq\sqrt{\frac{\beta}{2}}\|\bx\| \}},
\end{align*}
}
and 
the last inequality holds with probability exceeding $1-e^{\Omega(m\epsilon^2)}$ provided $m\gtrsim \epsilon^{-2}\log\epsilon^{-1}\cdot n$ (see Lemma~\ref{lem:aux_lemma2}).
\subsection{Proof of Lemma~\ref{lem:tech_lemma2}}
It follows from Lemma~\ref{lem:aux_lemma1} that $\frac{1}{2}\|\bx\|^2\leq\frac{1}{m}\|\by\|_1$ holds with probability $1-e^{\Omega(m)}$ provided $m\gtrsim n$. Thus, on the same event, we have 
\begin{align*}
&\ln\frac{1}{m}\sum_{k=1}^m(\ba_k^\top\bz)^s(\ba_k^\top\bx)^t g\lb\frac{|\ba_k^\top\bz|^2}{\|\bz\|^2}\rb h\lb\frac{m|\ba_k^\top\bx|^2}{\|\by\|_1}\rb\ba_k\ba_k^\top\rn\\
&\leq \ln\frac{1}{m}\sum_{k=1}^m |\ba_k^\top\bz|^s|\ba_k^\top\bx|^t g\lb\frac{|\ba_k^\top\bz|^2}{\|\bz\|^2}\rb\dsone{\beta\|\bz\|^2<|\ba_k^\top\bz|^2<\gamma\|\bz\|^2} \cdot\dsone{|\ba_k^\top\bx|^2<2\gamma\|\bx\|^2}\ba_k\ba_k^\top  \rn \\
&\leq\|\bz\|^s\|\bx\|^t\cdot 2^{\frac{t}{2}}\gamma^{\frac{s+t}{2}}\ln\frac{1}{m}\sum_{k=1}^m \dsone{|\ba_k^\top\bz|>\sqrt{\beta}\|\bz\|}\ba_k\ba_k^\top\rn \\
&\lesssim \|\bz\|^s\|\bx\|^t\cdot2^{\frac{t}{2}}\gamma^{\frac{s+t}{2}}\lb\sqrt{\beta}e^{-0.49\beta}+\epsilon\rb,
\end{align*}
where the last inequality holds with probability exceeding $1-e^{\Omega(m\epsilon^2)}$ provided $m\gtrsim \epsilon^{-2}\log\epsilon^{-1}\cdot n$; see Lemma~\ref{lem:aux_lemma2}.
\subsection{Proof of Lemma~\ref{lem:tech_lemma3}}
{ Firstly, similar to the proof of Lemma~\ref{lem:tech_lemma2},} $\frac{1}{2}\|\bx\|^2\leq\frac{1}{m}\|\by\|_1$ holds with probability $1-e^{\Omega(m)}$ provided $m\gtrsim n$. Thus, we have 
\begin{align*}
&\ln\frac{1}{m}\sum_{k=1}^m(\ba_k^\top\bz)^s(\ba_k^\top\bx)^t g\lb\frac{|\ba_k^\top\bz|^2}{\|\bz\|^2}\rb h\lb\frac{m|\ba_k^\top\bx|^2}{\|\by\|_1}\rb\bz\ba_k^\top\rn\\
&\leq \max_{\|\bu\|=\|\bv\|=1}\frac{1}{m}\sum_{k=1}^m |\ba_k^\top\bz|^s|\ba_k^\top\bx|^t g\lb\frac{|\ba_k^\top\bz|^2}{\|\bz\|^2}\rb\mathbf{1}_{\{\beta\|\bz\|^2<|\ba_k^\top\bz|<\gamma\|\bz\|^2\}}\cdot \dsone{|\ba_k^\top\bx|^2<2\gamma\|\bx\|^2}\cdot|\bm{u}^\top\bz|\cdot|\ba_k^\top\bm{v}|\\
&\leq \|\bz\|^{s+1}\|\bx\|^t\cdot 2^{\frac{t}{2}}\
\gamma^{\frac{s+t}{2}}\cdot \max_{\|\bm{v}\|=1}\frac{1}{m}\sum_{k=1}^m\dsone{|\ba_k^\top\bz|>\sqrt{\beta}\|\bz\|}|\ba_k^\top\bm{v}| \\
&\leq \|\bz\|^{s+1}\|\bx\|^t\cdot 2^{\frac{t}{2}}\
\gamma^{\frac{s+t}{2}}\cdot\max_{\|\bv\|=1} \sqrt{\frac{1}{m}\sum_{k=1}^m\dsone{|\ba_k^\top\bz|>\sqrt{\beta}\|\bz\|}|\ba_k^\top\bm{v}|^2}\\
& { \leq\|\bz\|^{s+1}\|\bx\|^t\cdot 2^{\frac{t}{2}}\
\gamma^{\frac{s+t}{2}}\sqrt{\sqrt{\beta}e^{-0.49\beta}+\epsilon}}\\
&\lesssim  \|\bz\|^{s+1}\|\bx\|^t\cdot 2^{\frac{t}{2}}\
\gamma^{\frac{s+t+1}{2}}\cdot \lb e^{-0.245\beta}+\sqrt{\epsilon}\rb,
\end{align*}
where the fourth inequality holds with probability exceeding $1-e^{\Omega(m\epsilon^2)}$ provided $m\gtrsim \epsilon^{-2}\log\epsilon^{-1}\cdot n$ which follows from Lemma~\ref{lem:aux_lemma2},
{ 
\begin{align*}
\frac{1}{m}\sum_{k=1}^m\dsone{|\ba_k^\top\bz|>\sqrt{\beta}\|\bz\|}|\ba_k^\top\bm{v}|^2 &=\lab\bv^\top\lb \frac{1}{m}\sum_{k=1}^m\dsone{|\ba_k^\top\bz|>\sqrt{\beta}\|\bz\|}\ba_k\ba_k^\top\rb\bv\rab\\
&\leq \ln \frac{1}{m}\sum_{k=1}^m\dsone{|\ba_k^\top\bz|>\sqrt{\beta}\|\bz\|}\ba_k\ba_k^\top\rn\\
&\leq \sqrt{\beta}e^{-0.49\beta}+\epsilon,
\end{align*}
}  
 {  and the last inequality follows from the fact $1<\beta<\gamma$ and $\sqrt{a+b}\leq\sqrt{a}+\sqrt{b}$}. 
\subsection{Proof of Lemma~\ref{lem:tech_lemma4}}
{ It follows from Lemma~\ref{lem:aux_lemma1} that $\frac{1}{2}\|\bx\|^2\leq\frac{1}{m}\|\by\|_1$ holds with probability $1-e^{\Omega(m)}$ provided $m\gtrsim n$}. Then a  simple algebra yields that 
\begin{align*}
&\lab\frac{1}{m}\sum_{k=1}^m(\ba_k^\top\bz)^s(\ba_k^\top\bx)^t g\lb\frac{|\ba_k^\top\bz|^2}{\|\bz\|^2}\rb h\lb\frac{m|\ba_k^\top\bx|^2}{\|\by\|_1}\rb \rab \\
&\leq \frac{1}{m}\sum_{k=1}^m|\ba_k^\top\bz|^s|\ba_k^\top\bx|^t g\lb\frac{|\ba_k^\top\bz|^2}{\|\bz\|^2}\rb h\lb\frac{m|\ba_k^\top\bx|^2}{\|\by\|_1}\rb\\
&\leq \frac{1}{m}\sum_{k=1}^m|\ba_k^\top\bz|^s|\ba_k^\top\bx|^t \dsone{\beta\|\bz\|^2\leq |\ba_k^\top\bz|^2\leq \gamma\|\bz\|^2} \dsone{|\ba_k^\top\bx|^2<2\gamma\|\bx\|^2}\\
&\leq \gamma^{\frac{s-2}{2}}\|\bz\|^{s-2}\cdot 2^{\frac{t}{2}}\gamma^{\frac{t}{2}}\|\bx\|^t\frac{1}{m}\sum_{k=1}^m|\ba_k^\top\bz|^2\dsone{|\ba_k^\top\bz|>\sqrt{\beta}\|\bz\|}\\
&\lesssim \|\bz\|^s\|\bx\|^t\cdot 2^{\frac{t}{2}}\gamma^{\frac{s+t}{2}}\lb\sqrt{\beta}e^{-0.49\beta}+\epsilon\rb,
\end{align*}
where the last inequality holds with probability exceeding $1-e^{\Omega(m\epsilon^2)}$ provided $m\gtrsim \epsilon^{-2}\log\epsilon^{-1}\cdot n$; see Lemma~\ref{lem:aux_lemma2}.
\section{Conclusion and outlook}\label{sec:conclusion}
A new loss function has been constructed for solving random systems of quadratic equations, which does not have  spurious local minima when the sampling complexity is optimal. This paper has focused on the real-valued  problem, and we will leave the examination of the complex case to future work. For the complex case, it is interesting to see whether the same loss function is still well-behaved under the optimal sampling complexity, or a more delicate activation function should be adopted.  
In addition, the technique presented in this paper may apply equally to the problem of reconstructing a general low rank matrix from   symmetric rank-$1$ projections \cite{CCGold15,KRTersti14,WWSS15}.

As stated at the beginning of this paper, the problem of solving systems of quadratic equations can be cast as a rank-$1$ matrix recovery problem. To see this, let $\A$ be a linear operator from $n\times n$ symmetric matrices to vectors of length $m$, defined as \begin{align*}
\A(\BZ) = \lcb\langle\BZ,\ba_k\ba_k^\top\rangle\rcb_{k=1}^m,\quad\forall~\BZ\in\R^{n\times n}\mbox{ being symmetric}.\numberthis\label{eq:A}
\end{align*}
Then a simple algebra yields that 
\begin{align*}
y_k = |\ba^\top_k\bx|^2 = \langle\ba_k\ba_k^\top,\BX\rangle,
\end{align*}
where $\BX=\bx\bx^\top$ is the  lift matrix defined associated with $\bx$.
Noticing the one to one correspondence between $\BX$ and $\bx$, instead of reconstructing $\bx$, one can attempt to reconstruct $\BX$ by seeking a rank-$1$ positive semidefinite matrix which fits the measurements as well as possible: 
\begin{align*}
\min_{\BZ}\frac{1}{2}\|\A(\BZ)-\by\|^2\quad\mbox{subject to}\quad \rank(\BZ)=1\mbox{ and }\BZ\succeq 0.\numberthis\label{eq:low_rank}
\end{align*}
Note that the  geometric landscape analysis presented in this paper as well as that in \cite{SQW:FCM:18} are carried out in the vector space. Instead, one can consider the geometric landscape of the loss function $\frac{1}{2}\|\A(\BZ)-\by\|^2$ on the embedded manifold of positive semidefinite rank-$1$ (or general rank-$r$) matrices under the rank-$1$ measurements. Moreover, it is worth studying  whether there exists a loss function on the lift matrix space which is well-behaved under the condition of optimal sampling complexity.
\appendix
\section{Auxiliary lemmas}
\begin{lemma}[\cite{Ver2011rand,CSV:CPAM:13}]\label{lem:aux_lemma1}
Assume $\ba_k\sim\N(0,\BI_n)$, $k=1,\cdots,m$, are independent. Then
\begin{align*}
\frac{1}{2}\leq\ln\frac{1}{m}\sum_{k=1}^m\ba_k\ba_k^\top\rn\leq 2
\end{align*}
hold with probability at least $1-2e^{-\Omega(m)}$ provided $m\gtrsim n$.
\end{lemma}
\begin{lemma}[\cite{CaiWeiphase}]\label{lem:aux_lemma2}
Fix $\eta\geq 1$ and let $\epsilon\in(0,1)$ be a sufficiently small constant. Assume $\ba_k\sim\N(0,\BI_n)$, $k=1,\cdots,m$, are independent. Then 
\begin{align*}
\ln\frac{1}{m}\sum_{k=1}^m\ba_k\ba_k^\top\dsone{|\ba_k^\top\bz|>\eta\|\bz\|}\rn
\lesssim \eta e^{-0.49\eta^2}+\epsilon
\end{align*}
holds uniformly for all $\|\bz\|\neq 0$ with probability exceeding $1-2e^{-\Omega(m\epsilon^2)}$ provided $m\gtrsim \epsilon^{-2}\log\epsilon^{-1}\cdot n$.
\end{lemma}

\section{Gradient and Hessian of the loss function}\label{app:gradHess}
Recall that 
\begin{align*}
f(\bz)=\frac{1}{2m}\sum_{k=1}^m\lb(\ba_k^\top\bz)^2-(\ba_k^\top\bx)^2\rb^2h\lb\frac{|\ba_k^\top\bz|^2}{\|\bz\|^2}\rb h\lb\frac{m|\ba_k^\top\bx|^2}{\|\by\|_1}\rb.
\end{align*}
By the chain rule we have 
\begin{align*}
\grad f(\bz) &= \frac{1}{m}\sum_{k=1}^m2\lb(\ba_k^\top\bz)^2-(\ba_k^\top\bx)^2\rb h\lb\frac{|\ba_k^\top\bz|^2}{\|\bz\|^2}\rb h\lb\frac{m|\ba_k^\top\bx|^2}{\|\by\|_1}\rb\ba_k\ba_k^\top\bz\\
&+\frac{1}{m}\sum_{k=1}^m\lb(\ba_k^\top\bz)^2-(\ba_k^\top\bx)^2\rb^2h'\lb\frac{|\ba_k^\top\bz|^2}{\|\bz\|^2}\rb h\lb\frac{m|\ba_k^\top\bx|^2}{\|\by\|_1}\rb\frac{\ba_k\ba_k^\top\bz}{\|\bz\|^2}\\
&-\frac{1}{m}\sum_{k=1}^m\lb(\ba_k^\top\bz)^2-(\ba_k^\top\bx)^2\rb^2h'\lb\frac{|\ba_k^\top\bz|^2}{\|\bz\|^2}\rb h\lb\frac{m|\ba_k^\top\bx|^2}{\|\by\|_1}\rb\frac{(\ba_k^\top\bz)^2\bz}{\|\bz\|^4}.
\end{align*}
In order to compute $\hessian f(\bz)$, let 
\begin{align*}
g_{1k} &= 2\lb(\ba_k^\top\bz)^2-(\ba_k^\top\bx)^2\rb h\lb\frac{|\ba_k^\top\bz|^2}{\|\bz\|^2}\rb h\lb\frac{m|\ba_k^\top\bx|^2}{\|\by\|_1}\rb\ba_k\ba_k^\top\bz,\\
g_{2k} & = \lb(\ba_k^\top\bz)^2-(\ba_k^\top\bx)^2\rb^2h'\lb\frac{|\ba_k^\top\bz|^2}{\|\bz\|^2}\rb h\lb\frac{m|\ba_k^\top\bx|^2}{\|\by\|_1}\rb\frac{\ba_k\ba_k^\top\bz}{\|\bz\|^2},\\
g_{3k} & = -\lb(\ba_k^\top\bz)^2-(\ba_k^\top\bx)^2\rb^2h'\lb\frac{|\ba_k^\top\bz|^2}{\|\bz\|^2}\rb h\lb\frac{m|\ba_k^\top\bx|^2}{\|\by\|_1}\rb\frac{(\ba_k^\top\bz)^2\bz}{\|\bz\|^4}.
\end{align*}
Then we have 
\begin{align*}
\hessian f(\bz) = \frac{1}{m}\sum_{k=1}^m J_{g_{1k}} +  J_{g_{2k}}+ J_{g_{3k}},\numberthis\label{eq:hessian}
\end{align*}
where $J_{g_{1k}}$, $J_{g_{2k}}$ and $J_{g_{3k}}$ are the Jacobian matrices of $g_{1k}$, $g_{2k}$ and $g_{3k}$ respectively, given by 
\begin{align*}
J_{g_{1k}} & = 2\lb3(\ba_k^\top\bz)^2-(\ba_k^\top\bx)^2\rb h\lb\frac{|\ba_k^\top\bz|^2}{\|\bz\|^2}\rb h\lb\frac{m|\ba_k^\top\bx|^2}{\|\by\|_1}\rb\ba_k\ba_k^\top\\
&+\frac{4}{\|\bz\|^2}\lb (\ba_k^\top\bz)^2-(\ba_k^\top\bx)^2\rb(\ba_k^\top\bz)^2h'\lb\frac{|\ba_k^\top\bz|^2}{\|\bz\|^2}\rb h\lb\frac{m|\ba_k^\top\bx|^2}{\|\by\|_1}\rb\ba_k\ba_k^\top\\
&-\frac{4}{\|\bz\|^4}\lb (\ba_k^\top\bz)^2-(\ba_k^\top\bx)^2\rb(\ba_k^\top\bz)^3h'\lb\frac{|\ba_k^\top\bz|^2}{\|\bz\|^2}\rb h\lb\frac{m|\ba_k^\top\bx|^2}{\|\by\|_1}\rb\ba_k\bz^\top,\\
\\
J_{g_{2k}} & =\frac{1}{\|\bz\|^2}\lb 5(\ba_k^\top\bz)^4-6(\ba_k^\top\bz)^2(\ba_k^\top\bx)^2+(\ba_k^\top\bx)^4\rb h'\lb\frac{|\ba_k^\top\bz|^2}{\|\bz\|^2}\rb h\lb\frac{m|\ba_k^\top\bx|^2}{\|\by\|_1}\rb\ba_k\ba_k^\top\\
&+\frac{2}{\|\bz\|^4} \lb(\ba_k^\top\bz)^2-(\ba_k^\top\bx)^2\rb^2
(\ba_k^\top\bz)^2h''\lb\frac{|\ba_k^\top\bz|^2}{\|\bz\|^2}\rb h\lb\frac{m|\ba_k^\top\bx|^2}{\|\by\|_1}\rb\ba_k\ba_k^\top\\
&-\frac{2}{\|\bz\|^6} \lb(\ba_k^\top\bz)^2-(\ba_k^\top\bx)^2\rb^2
(\ba_k^\top\bz)^3h''\lb\frac{|\ba_k^\top\bz|^2}{\|\bz\|^2}\rb h\lb\frac{m|\ba_k^\top\bx|^2}{\|\by\|_1}\rb\ba_k\bz^\top\\
&-\frac{2}{\|\bz\|^4}\lb(\ba_k^\top\bz)^2-(\ba_k^\top\bx)^2\rb^2(\ba_k^\top\bz)h'\lb\frac{|\ba_k^\top\bz|^2}{\|\bz\|^2}\rb h\lb\frac{m|\ba_k^\top\bx|^2}{\|\by\|_1}\rb\ba_k\bz^\top,\\
\\
J_{g_{3k}} &=-\frac{1}{\|\bz\|^4}\lb6(\ba_k^\top\bz)^5-8(\ba_k^\top\bz)^3(\ba_k^\top\bx)^2+2(\ba_k^\top\bz)(\ba_k^\top\bx)^4\rb h'\lb\frac{|\ba_k^\top\bz|^2}{\|\bz\|^2}\rb h\lb\frac{m|\ba_k^\top\bx|^2}{\|\by\|_1}\rb\bz\ba_k^\top\\
&-\frac{2}{\|\bz\|^6}\lb(\ba_k^\top\bz)^2-(\ba_k^\top\bx)^2\rb^2(\ba_k^\top\bz)^3h''\lb\frac{|\ba_k^\top\bz|^2}{\|\bz\|^2}\rb h\lb\frac{m|\ba_k^\top\bx|^2}{\|\by\|_1}\rb\bz\ba_k^\top\\
&+\frac{2}{\|\bz\|^8}\lb(\ba_k^\top\bz)^2-(\ba_k^\top\bx)^2\rb^2(\ba_k^\top\bz)^4h''\lb\frac{|\ba_k^\top\bz|^2}{\|\bz\|^2}\rb h\lb\frac{m|\ba_k^\top\bx|^2}{\|\by\|_1}\rb\bz\bz^\top\\
&+\frac{4}{\|\bz\|^6}\lb(\ba_k^\top\bz)^2-(\ba_k^\top\bx)^2\rb^2(\ba_k^\top\bz)^2h'\lb\frac{|\ba_k^\top\bz|^2}{\|\bz\|^2}\rb h\lb\frac{m|\ba_k^\top\bx|^2}{\|\by\|_1}\rb\bz\bz^\top\\
&-\frac{1}{\|\bz\|^4}\lb(\ba_k^\top\bz)^2-(\ba_k^\top\bx)^2\rb^2h'\lb\frac{|\ba_k^\top\bz|^2}{\|\bz\|^2}\rb h\lb\frac{m|\ba_k^\top\bx|^2}{\|\by\|_1}\rb(\ba_k^\top\bz)^2\BI.
\end{align*}
It is worth noting that even though each Jacobian matrix is not symmetric, their sum is indeed symmetric which satisfies the symmetric property of a Hessian matrix. To see this, adding all the terms involving $\ba_k\bz^\top$ and $\bz\ba_k^\top$ together gives
\begin{align*}
&-\frac{2}{\|\bz\|^6} \lb(\ba_k^\top\bz)^2-(\ba_k^\top\bx)^2\rb^2
(\ba_k^\top\bz)^3h''\lb\frac{|\ba_k^\top\bz|^2}{\|\bz\|^2}\rb h\lb\frac{m|\ba_k^\top\bx|^2}{\|\by\|_1}\rb(\ba_k\bz^\top+\bz\ba_k^\top)\\
&-\frac{1}{\|\bz\|^4}\lb6(\ba_k^\top\bz)^5-8(\ba_k^\top\bz)^3(\ba_k^\top\bx)^2+2(\ba_k^\top\bz)(\ba_k^\top\bx)^4\rb h'\lb\frac{|\ba_k^\top\bz|^2}{\|\bz\|^2}\rb h\lb\frac{m|\ba_k^\top\bx|^2}{\|\by\|_1}\rb(\ba_k^\top\bz+\bz\ba_k^\top).
\end{align*}

{To further check the correctness of our calculations, we consider the special case when  $n=1$, and compare  the values of $f'(z)$ and $f''(z)$ computed using the derived formulas with that computed via the following  finite difference schemes:
\begin{align*}
f'(z) \approx \frac{f(z+\epsilon)-f(z-\epsilon)}{2\epsilon}\quad\mbox{and}\quad f''(z)\approx\frac{f(z+\epsilon)-2f(z)+f(z-\epsilon)}{\epsilon^2},
\end{align*}
where $\epsilon>0$ is a small constant (here we choose $\epsilon=10^{-5}$). Table~\ref{table:GH} includes the computational results for the fixed $x=1$ and a few randomly generated $z$.

}
\begin{table}[t!]
\centering
\caption{   Computational results from computing $f'(z)$ and $f''(z)$ via the formulas and the finite difference schemes. Here, $x=1$, $\{a_k\}_{k=1}^m$ ($m=128$) is a set of standard Gaussian random variables, and the results for three $z$'s in {$\{4.3042, 1.7588, 0.5544
\}$} are presented. Each $z$  is uniformly sampled from $[0,10]$ .}
\label{table:GH}
\makegapedcells
\setcellgapes{3pt}
\begin{tabular}{c|ccc|ccc}
\hline
 & \multicolumn{3}{c|}{$f'(z)$ for three $z$'s} & \multicolumn{3}{c}{$f''(z)$ for three $z$'s}\\
 \hline 
 By formulas & 787.0769 & 38.4171 & -4.0066 & 569.4601 &86.3959 &-0.8143 \\
 \hline
 By finite difference & 787.0769 & 38.4171 & -4.0066 & 569.4601 &86.3959 &-0.8143 \\
 \hline
\end{tabular}
\end{table}

\subsection{Proof of \eqref{eq:thmR1_eq2a0}}\label{sec:app:sub3}
Denote by $B_i,~i=2,\cdots, 12$ the spectral norm of the $i$-th matrix in the Hessian expression \eqref{eq:hessian}. 
\paragraph{{Bound for $B_2$}}
\begin{align*}
B_2&\leq \frac{4}{\|\bz\|^2}\left\|\frac{1}{m}\sum_{k=1}^m \left[ (\ba_k^\top\bz)^4-(\ba_k^\top\bz)^2(\ba_k^\top\bx)^2\right]h'\left(\frac{|\ba_k^\top\bz|^2}{\|\bz\|^2}\right)h\left(\frac{m|\ba_k^\top\bx|^2}{\|\by\|_1}\right)\ba_k\ba_k^\top\right\|\\
&\leq \frac{4}{\|\bz\|^2}\left\|\frac{1}{m}\sum_{k=1}^m  (\ba_k^\top\bz)^4h'\left(\frac{|\ba_k^\top\bz|^2}{\|\bz\|^2}\right)h\left(\frac{m|\ba_k^\top\bx|^2}{\|\by\|_1}\right)\ba_k\ba_k^\top\right\|\\
&+\frac{4}{\|\bz\|^2}\frac{1}{m}\sum_{k=1}^m \left[ (\ba_k^\top\bz)^2(\ba_k^\top\bx)^2h'\left(\frac{|\ba_k^\top\bz|^2}{\|\bz\|^2}\right)h\left(\frac{m|\ba_k^\top\bx|^2}{\|\by\|_1}\right)\ba_k\ba_k^\top\right\|\\
&\lesssim \frac{4}{\|\bz\|^2}\cdot\|\bz\|^4\cdot |h'|_{\infty}\cdot\gamma^2\left(\sqrt{\beta}e^{-0.49\beta}+\epsilon\right)+\frac{4}{\|\bz\|^2}\cdot\|\bz\|^2\|\bx\|^2\cdot |h'|_{\infty}\cdot 2\gamma^2\left(\sqrt{\beta}e^{-0.49\beta}+\epsilon\right)\\
    &\lesssim \gamma^2\left(\sqrt{\beta}e^{-0.49\beta}+\epsilon\right)\cdot|h'|_{\infty}\max\left\{\|\bz\|^2,\|\bx\|^2 \right\},
\end{align*}
where the third inequality follows from Lemma \ref{lem:tech_lemma2} with $(s,t)=(4,0), (2,2)$, respectively.
\paragraph{{Bound for $B_3$}}
\begin{align*}
B_3 &\leq \frac{4}{\|\bz\|^4}\left\|\frac{1}{m}\sum_{k=1}^m \left[ (\ba_k^\top\bz)^5-(\ba_k^\top\bz)^3(\ba_k^\top\bx)^2\right]h'\left(\frac{|\ba_k^\top\bz|^2}{\|\bz\|^2}\right)h\left(\frac{m|\ba_k^\top\bx|^2}{\|\by\|_1}\right)\ba_k\bz^\top \right\|\\
&\leq  \frac{4}{\|\bz\|^4}\left\|\frac{1}{m}\sum_{k=1}^m  (\ba_k^\top\bz)^5h'\left(\frac{|\ba_k^\top\bz|^2}{\|\bz\|^2}\right)h\left(\frac{m|\ba_k^\top\bx|^2}{\|\by\|_1}\right)\ba_k\bz^\top \right\|\\
&+ \frac{4}{\|\bz\|^4}\left\|\frac{1}{m}\sum_{k=1}^m (\ba_k^\top\bz)^3(\ba_k^\top\bx)^2h'\left(\frac{|\ba_k^\top\bz|^2}{\|\bz\|^2}\right)h\left(\frac{m|\ba_k^\top\bx|^2}{\|\by\|_1}\right)\ba_k\bz^\top \right\|\\
    &\lesssim \frac{4}{\|\bz\|^4}\cdot \|\bz\|^6\cdot |h'|_{\infty}\cdot\gamma^3\left(e^{-0.245\beta}+\sqrt{\epsilon}\right)+\frac{4}{\|\bz\|^4}\cdot \|\bz\|^4\|\bx\|^2\cdot |h'|_{\infty}\cdot 2\gamma^3\left(e^{-0.245\beta}+\sqrt{\epsilon}\right)\\
    &\lesssim \gamma^3\left(e^{-0.245\beta}+\sqrt{\epsilon}\right)\cdot|h'|_{\infty}\max\left\{\|\bz\|^2,\|\bx\|^2 \right\},
\end{align*}
where the third inequality follows from Lemma \ref{lem:tech_lemma3} with $(s,t)=(5,0), (3,2)$, respectively.
\paragraph{Bound for $B_4$}
\begin{align*}B_4&\leq 
\frac{1}{\|\bz\|^2}\left\|\frac{1}{m}\sum_{k=1}^m\left[ 5(\ba_k^\top\bz)^4-6(\ba_k^\top\bz)^2(\ba_k^\top\bx)^2+(\ba_k^\top\bx)^4\right]h'\left(\frac{|\ba_k^\top\bz|^2}{\|\bz\|^2}\right)h\left(\frac{m|\ba_k^\top\bx|^2}{\|\by\|_1}\right)\ba_k\ba_k^\top \right\|\\
&\leq \frac{1}{\|\bz\|^2}\left\|\frac{1}{m}\sum_{k=1}^m 5(\ba_k^\top\bz)^4h'\left(\frac{|\ba_k^\top\bz|^2}{\|\bz\|^2}\right)h\left(\frac{m|\ba_k^\top\bx|^2}{\|\by\|_1}\right)\ba_k\ba_k^\top \right\|\\
& +\frac{1}{\|\bz\|^2}\left\|\frac{1}{m}\sum_{k=1}^m6(\ba_k^\top\bz)^2(\ba_k^\top\bx)^2h'\left(\frac{|\ba_k^\top\bz|^2}{\|\bz\|^2}\right)h\left(\frac{m|\ba_k^\top\bx|^2}{\|\by\|_1}\right)\ba_k\ba_k^\top \right\|\\
&+\frac{1}{\|\bz\|^2}\left\|\frac{1}{m}\sum_{k=1}^m(\ba_k^\top\bx)^4h'\left(\frac{|\ba_k^\top\bz|^2}{\|\bz\|^2}\right)h\left(\frac{m|\ba_k^\top\bx|^2}{\|\by\|_1}\right)\ba_k\ba_k^\top \right\|\\
    &\lesssim \frac{1}{\|\bz\|^2}\cdot 5\|\bz\|^4\cdot|h'|_{\infty}\cdot \gamma^2\left(\sqrt{\beta}e^{-0.49\beta}+\epsilon\right) + \frac{1}{\|\bz\|^2}\cdot 6\|\bz\|^2\|\bx\|^2\cdot|h'|_{\infty}\cdot 2\gamma^2\left(\sqrt{\beta}e^{-0.49\beta}+\epsilon\right)\\
    &+ \frac{1}{\|\bz\|^2}\cdot \|\bx\|^4\cdot|h'|_{\infty}\cdot 4\gamma^2\left(\sqrt{\beta}e^{-0.49\beta}+\epsilon\right)\\
    &\lesssim \gamma^2\left(\sqrt{\beta}e^{-0.49\beta}+\epsilon\right)\cdot |h'|_{\infty}\max\left\{\|\bz\|^2,\|\bx\|^2,\frac{\|\bx\|^4}{\|\bz\|^2} \right\},
\end{align*}
where the third inequality follows from Lemma \ref{lem:tech_lemma2} with $(s,t)=(4,0), (2,2), (0,4)$, respectively.
\paragraph{Bound for $B_5$}
\begin{align*}
B_5&\leq \frac{2}{\|\bz\|^4}\left\|\frac{1}{m}\sum_{k=1}^m\left[ (\ba_k^\top\bz)^6-2(\ba_k^\top\bz)^4(\ba_k^\top\bx)^2+(\ba_k^\top\bz)^2(\ba_k^\top\bx)^4\right]h''\left(\frac{|\ba_k^\top\bz|^2}{\|\bz\|^2}\right)h\left(\frac{m|\ba_k^\top\bx|^2}{\|\by\|_1}\right)\ba_k\ba_k^\top \right\|\\
&\leq \frac{2}{\|\bz\|^4}\left\|\frac{1}{m}\sum_{k=1}^m(\ba_k^\top\bz)^6h''\left(\frac{|\ba_k^\top\bz|^2}{\|\bz\|^2}\right)h\left(\frac{m|\ba_k^\top\bx|^2}{\|\by\|_1}\right)\ba_k\ba_k^\top \right\|\\
&+\frac{2}{\|\bz\|^4}\left\|\frac{1}{m}\sum_{k=1}^m2(\ba_k^\top\bz)^4(\ba_k^\top\bx)^2h''\left(\frac{|\ba_k^\top\bz|^2}{\|\bz\|^2}\right)h\left(\frac{m|\ba_k^\top\bx|^2}{\|\by\|_1}\right)\ba_k\ba_k^\top \right\|\\
&+\frac{2}{\|\bz\|^4}\left\|\frac{1}{m}\sum_{k=1}^m(\ba_k^\top\bz)^2(\ba_k^\top\bx)^4h''\left(\frac{|\ba_k^\top\bz|^2}{\|\bz\|^2}\right)h\left(\frac{m|\ba_k^\top\bx|^2}{\|\by\|_1}\right)\ba_k\ba_k^\top \right\|\\
    &\lesssim \frac{2}{\|\bz\|^4}\cdot \|\bz\|^6\cdot |h''|_{\infty}\cdot \gamma^3 \left(\sqrt{\beta}e^{-0.49\beta}+\epsilon\right)+\frac{4}{\|\bz\|^4}\cdot \|\bz\|^4\|\bx\|^2\cdot |h''|_{\infty}\cdot 2\gamma^3 \left(\sqrt{\beta}e^{-0.49\beta}+\epsilon\right)\\
    &+\frac{2}{\|\bz\|^4}\cdot \|\bz\|^2\|\bx\|^4\cdot |h''|_{\infty}\cdot 4\gamma^3 \left(\sqrt{\beta}e^{-0.49\beta}+\epsilon\right)\\
    &\lesssim \gamma^3 \left(\sqrt{\beta}e^{-0.49\beta}+\epsilon\right)\cdot |h''|_{\infty}\max\left\{\|\bz\|^2,\|\bx\|^2,\frac{\|\bx\|^4}{\|\bz\|^2} \right\},
\end{align*}
where the third inequality follows from Lemma \ref{lem:tech_lemma2}, with $(s,t)=(6,0), (4,2), (2,4)$, respectively.

\paragraph{Bound for $B_6$}
\begin{align*}
B_6&\leq \frac{2}{\|\bz\|^6}\left\|\frac{1}{m}\sum_{k=1}^m\left[ (\ba_k^\top\bz)^7-2(\ba_k^\top\bz)^5(\ba_k^\top\bx)^2+(\ba_k^\top\bz)^3(\ba_k^\top\bx)^4\right]h''\left(\frac{|\ba_k^\top\bz|^2}{\|\bz\|^2}\right)h\left(\frac{m|\ba_k^\top\bx|^2}{\|\by\|_1}\right)\ba_k\bz^\top \right\|\\
&\frac{2}{\|\bz\|^6}\left\|\frac{1}{m}\sum_{k=1}^m (\ba_k^\top\bz)^7h''\left(\frac{|\ba_k^\top\bz|^2}{\|\bz\|^2}\right)h\left(\frac{m|\ba_k^\top\bx|^2}{\|\by\|_1}\right)\ba_k\bz^\top \right\|\\
&+\frac{2}{\|\bz\|^6}\left\|\frac{1}{m}\sum_{k=1}^m2(\ba_k^\top\bz)^5(\ba_k^\top\bx)^2h''\left(\frac{|\ba_k^\top\bz|^2}{\|\bz\|^2}\right)h\left(\frac{m|\ba_k^\top\bx|^2}{\|\by\|_1}\right)\ba_k\bz^\top \right\|\\
&+\frac{2}{\|\bz\|^6}\left\|\frac{1}{m}\sum_{k=1}^m(\ba_k^\top\bz)^3(\ba_k^\top\bx)^4h''\left(\frac{|\ba_k^\top\bz|^2}{\|\bz\|^2}\right)h\left(\frac{m|\ba_k^\top\bx|^2}{\|\by\|_1}\right)\ba_k\bz^\top \right\|\\
    &\lesssim \frac{2}{\|\bz\|^6}\cdot \|\bz\|^8\cdot |h''|_{\infty}\cdot\gamma^4\left( e^{-0.245\beta}+\sqrt{\epsilon}\right)+\frac{4}{\|\bz\|^6}\cdot \|\bz\|^6\|\bx\|^2\cdot |h''|_{\infty}\cdot2\gamma^4\left( e^{-0.245\beta}+\sqrt{\epsilon}\right)\\
    &+\frac{2}{\|\bz\|^6}\cdot \|\bz\|^4\|\bx\|^4\cdot |h''|_{\infty}\cdot4\gamma^4\left( e^{-0.245\beta}+\sqrt{\epsilon}\right)\\
    &\lesssim  \gamma^4\left( e^{-0.245\beta}+\sqrt{\epsilon}\right)\cdot |h''|_{\infty}\max\left\{\|\bz\|^2,\|\bx\|^2,\frac{\|\bx\|^4}{\|\bz\|^2} \right\},
\end{align*}
where the third inequality follows from Lemma \ref{lem:tech_lemma3} with $(s,t)=(7,0), (5,2), (3,4)$, respectively.

\paragraph{Bound for $B_7$}
\begin{align*}
B_7&\leq \frac{2}{\|\bz\|^4} \left\|\frac{1}{m}\sum_{k=1}^m\left[ (\ba_k^\top\bz)^5-2(\ba_k^\top\bz)^3(\ba_k^\top\bx)^2+(\ba_k^\top\bz)(\ba_k^\top\bx)^4\right]h'\left(\frac{|\ba_k^\top\bz|^2}{\|\bz\|^2}\right)h\left(\frac{m|\ba_k^\top\bx|^2}{\|\by\|_1}\right)\ba_k\bz^\top \right\|\\
&\leq \frac{2}{\|\bz\|^4} \left\|\frac{1}{m}\sum_{k=1}^m(\ba_k^\top\bz)^5h'\left(\frac{|\ba_k^\top\bz|^2}{\|\bz\|^2}\right)h\left(\frac{m|\ba_k^\top\bx|^2}{\|\by\|_1}\right)\ba_k\bz^\top \right\|\\
&+\frac{2}{\|\bz\|^4} \left\|\frac{1}{m}\sum_{k=1}^m2(\ba_k^\top\bz)^3(\ba_k^\top\bx)^2h'\left(\frac{|\ba_k^\top\bz|^2}{\|\bz\|^2}\right)h\left(\frac{m|\ba_k^\top\bx|^2}{\|\by\|_1}\right)\ba_k\bz^\top \right\|\\
&+\frac{2}{\|\bz\|^4} \left\|\frac{1}{m}\sum_{k=1}^m(\ba_k^\top\bz)(\ba_k^\top\bx)^4h'\left(\frac{|\ba_k^\top\bz|^2}{\|\bz\|^2}\right)h\left(\frac{m|\ba_k^\top\bx|^2}{\|\by\|_1}\right)\ba_k\bz^\top \right\|\\
    &\lesssim \frac{2}{\|\bz\|^4}\cdot \|\bz\|^6\cdot |h'|_{\infty}\cdot\gamma^3 \left( e^{-0.245\beta}+\sqrt{\epsilon}\right)+\frac{4}{\|\bz\|^4}\cdot \|\bz\|^4\|\bx\|^2\cdot |h'|_{\infty}\cdot2\gamma^3 \left( e^{-0.245\beta}+\sqrt{\epsilon}\right)\\
    &+\frac{2}{\|\bz\|^4}\cdot \|\bz\|^2\|\bx\|^4\cdot |h'|_{\infty}\cdot4\gamma^3 \left( e^{-0.245\beta}+\sqrt{\epsilon}\right)\\
    &\lesssim \gamma^3 \left( e^{-0.245\beta}+\sqrt{\epsilon}\right)\cdot |h'|_{\infty}\max\left\{\|\bz\|^2,\|\bx\|^2,\frac{\|\bx\|^4}{\|\bz\|^2} \right\},
\end{align*}
where the third inequality follows from Lemma \ref{lem:tech_lemma3} with $(s,t)=(5,0), (3,2), (1,4)$, respectively.

\paragraph{Bound for $B_8$}
\begin{align*}
B_8&\leq\frac{1}{\|\bz\|^4}\left\|\frac{1}{m}\sum_{k=1}^m\left[ 6(\ba_k^\top\bz)^5-8(\ba_k^\top\bz)^3(\ba_k^\top\bx)^2+2(\ba_k^\top\bz)(\ba_k^\top\bx)^4\right]h'\left(\frac{|\ba_k^\top\bz|^2}{\|\bz\|^2}\right)h\left(\frac{m|\ba_k^\top\bx|^2}{\|\by\|_1}\right)\bz\ba_k^\top \right\|\\
&\leq \frac{1}{\|\bz\|^4}\left\|\frac{1}{m}\sum_{k=1}^m 6(\ba_k^\top\bz)^5h'\left(\frac{|\ba_k^\top\bz|^2}{\|\bz\|^2}\right)h\left(\frac{m|\ba_k^\top\bx|^2}{\|\by\|_1}\right)\bz\ba_k^\top \right\|\\
&+\frac{1}{\|\bz\|^4}\left\|\frac{1}{m}\sum_{k=1}^m8(\ba_k^\top\bz)^3(\ba_k^\top\bx)^2h'\left(\frac{|\ba_k^\top\bz|^2}{\|\bz\|^2}\right)h\left(\frac{m|\ba_k^\top\bx|^2}{\|\by\|_1}\right)\bz\ba_k^\top \right\|\\
&+\frac{1}{\|\bz\|^4}\left\|\frac{1}{m}\sum_{k=1}^m2(\ba_k^\top\bz)(\ba_k^\top\bx)^4h'\left(\frac{|\ba_k^\top\bz|^2}{\|\bz\|^2}\right)h\left(\frac{m|\ba_k^\top\bx|^2}{\|\by\|_1}\right)\bz\ba_k^\top \right\|\\
    &\lesssim \frac{1}{\|\bz\|^4}\cdot 6\|\bz\|^6\cdot |h'|_{\infty}\cdot \gamma^3\left(e^{-0.245\beta}+\sqrt{\epsilon}\right)+ \frac{1}{\|\bz\|^4}\cdot 8\|\bz\|^4\|\bx\|^2\cdot |h'|_{\infty}\cdot 2\gamma^3\left(e^{-0.245\beta}+\sqrt{\epsilon}\right)\\
    &+ \frac{1}{\|\bz\|^4}\cdot 2\|\bz\|^2\|\bx\|^4\cdot |h'|_{\infty}\cdot 4\gamma^3\left(e^{-0.245\beta}+\sqrt{\epsilon}\right)\\
    &\lesssim 30\gamma^3\left(e^{-0.245\beta}+\sqrt{\epsilon}\right)\cdot |h'|_{\infty}\max\left\{\|\bz\|^2,\|\bx\|^2,\frac{\|\bx\|^4}{\|\bz\|^2} \right\},
\end{align*}
where the third inequality follows from Lemma \ref{lem:tech_lemma3} with $(s,t)=(5,0), (3,2), (1,4)$, respectively. 

\paragraph{Bound for $B_9$}
\begin{align*}
B_9 & \leq \frac{2}{\|\bz\|^6}\left\|\frac{1}{m}\sum_{k=1}^m\left[ (\ba_k^\top\bz)^7-2(\ba_k^\top\bz)^5(\ba_k^\top\bx)^2+(\ba_k^\top\bz)^3(\ba_k^\top\bx)^4\right]h''\left(\frac{|\ba_k^\top\bz|^2}{\|\bz\|^2}\right)h\left(\frac{m|\ba_k^\top\bx|^2}{\|\by\|_1}\right)\bz\ba_k^\top \right\|\\
&\leq \frac{2}{\|\bz\|^6}\left\|\frac{1}{m}\sum_{k=1}^m (\ba_k^\top\bz)^7h''\left(\frac{|\ba_k^\top\bz|^2}{\|\bz\|^2}\right)h\left(\frac{m|\ba_k^\top\bx|^2}{\|\by\|_1}\right)\bz\ba_k^\top \right\|\\
&+\frac{2}{\|\bz\|^6}\left\|\frac{1}{m}\sum_{k=1}^m2(\ba_k^\top\bz)^5(\ba_k^\top\bx)^2h''\left(\frac{|\ba_k^\top\bz|^2}{\|\bz\|^2}\right)h\left(\frac{m|\ba_k^\top\bx|^2}{\|\by\|_1}\right)\bz\ba_k^\top \right\|\\
&+\frac{2}{\|\bz\|^6}\left\|\frac{1}{m}\sum_{k=1}^m(\ba_k^\top\bz)^3(\ba_k^\top\bx)^4h''\left(\frac{|\ba_k^\top\bz|^2}{\|\bz\|^2}\right)h\left(\frac{m|\ba_k^\top\bx|^2}{\|\by\|_1}\right)\bz\ba_k^\top \right\|\\
    &\lesssim \frac{2}{\|\bz\|^6}\cdot \|\bz\|^8\cdot |h''|_{\infty}\cdot \gamma^4\left(e^{-0.245\beta}+\sqrt{\epsilon}\right)+\frac{4}{\|\bz\|^6}\cdot \|\bz\|^6\|\bx\|^2\cdot |h''|_{\infty}\cdot 2\gamma^4\left(e^{-0.245\beta}+\sqrt{\epsilon}\right)\\
    &+\frac{2}{\|\bz\|^6}\cdot \|\bz\|^4\|\bx\|^4\cdot |h''|_{\infty}\cdot 4\gamma^4\left(e^{-0.245\beta}+\sqrt{\epsilon}\right)\\
    &\lesssim\gamma^4\left(e^{-0.245\beta}+\sqrt{\epsilon}\right)\cdot |h''|_{\infty}\max\left\{\|\bz\|^2,\|\bx\|^2,\frac{\|\bx\|^4}{\|\bz\|^2} \right\},
\end{align*}
where the third inequality follows from Lemma \ref{lem:tech_lemma3} with $(s,t)=(7,0), (5,2), (3,4)$, respectively.

\paragraph{Bound for $B_{10}$}
\begin{align*}
B_{10}&\leq \frac{2}{\|\bz\|^8}\left\|\frac{1}{m}\sum_{k=1}^m\left[ (\ba_k^\top\bz)^8-2(\ba_k^\top\bz)^6(\ba_k^\top\bx)^2+(\ba_k^\top\bz)^4(\ba_k^\top\bx)^4\right]h''\left(\frac{|\ba_k^\top\bz|^2}{\|\bz\|^2}\right)h\left(\frac{m|\ba_k^\top\bx|^2}{\|\by\|_1}\right)\bz\bz^\top \right\|\\
&\leq \frac{2}{\|\bz\|^8}\left\|\frac{1}{m}\sum_{k=1}^m(\ba_k^\top\bz)^8h''\left(\frac{|\ba_k^\top\bz|^2}{\|\bz\|^2}\right)h\left(\frac{m|\ba_k^\top\bx|^2}{\|\by\|_1}\right)\bz\bz^\top \right\|\\
&+\frac{2}{\|\bz\|^8}\left\|\frac{1}{m}\sum_{k=1}^m(\ba_k^\top\bz)^4(\ba_k^\top\bx)^4h''\left(\frac{|\ba_k^\top\bz|^2}{\|\bz\|^2}\right)h\left(\frac{m|\ba_k^\top\bx|^2}{\|\by\|_1}\right)\bz\bz^\top \right\|\\
&+\frac{2}{\|\bz\|^8}\left\|\frac{1}{m}\sum_{k=1}^m(\ba_k^\top\bz)^4(\ba_k^\top\bx)^4h''\left(\frac{|\ba_k^\top\bz|^2}{\|\bz\|^2}\right)h\left(\frac{m|\ba_k^\top\bx|^2}{\|\by\|_1}\right)\bz\bz^\top \right\|\\
    &\lesssim \frac{2}{\|\bz\|^8}\cdot \|\bz\|^{10}\cdot|h''|_{\infty}\cdot \gamma^4\left(\sqrt{\beta}e^{-0.49\beta}+\epsilon\right)+ \frac{4}{\|\bz\|^8}\cdot \|\bz\|^8\|\bx\|^2\cdot|h''|_{\infty}\cdot 2\gamma^4\left(\sqrt{\beta}e^{-0.49\beta}+\epsilon\right)\\
    &+ \frac{2}{\|\bz\|^8}\cdot \|\bz\|^6\|\bx\|^4\cdot|h''|_{\infty}\cdot 4\gamma^4\left(\sqrt{\beta}e^{-0.49\beta}+\epsilon\right)\\
    &\lesssim \gamma^4\left(\sqrt{\beta}e^{-0.49\beta}+\epsilon\right)\cdot |h''|_{\infty}\max\left\{\|\bz\|^2,\|\bx\|^2,\frac{\|\bx\|^4}{\|\bz\|^2} \right\},
\end{align*}
where the third inequality follows from  the fact $\|\bz\bz^\top\|=\|\bz\|^2$ and Lemma \ref{lem:tech_lemma4} with $(s,t)=(8,0),(6,2),(4,4)$, respectively.

\paragraph{Bound for $B_{11}$}
\begin{align*}
B_{11}&\leq \frac{4}{\|\bz\|^6}\left\|\frac{1}{m}\sum_{k=1}^m\left[ (\ba_k^\top\bz)^6-2(\ba_k^\top\bz)^4(\ba_k^\top\bx)^2+(\ba_k^\top\bz)^2(\ba_k^\top\bx)^4\right]h'\left(\frac{|\ba_k^\top\bz|^2}{\|\bz\|^2}\right)h\left(\frac{m|\ba_k^\top\bx|^2}{\|\by\|_1}\right)\bz\bz^\top \right\|\\
&\leq \frac{4}{\|\bz\|^6}\left\|\frac{1}{m}\sum_{k=1}^m (\ba_k^\top\bz)^6h'\left(\frac{|\ba_k^\top\bz|^2}{\|\bz\|^2}\right)h\left(\frac{m|\ba_k^\top\bx|^2}{\|\by\|_1}\right)\bz\bz^\top \right\|\\
&+\frac{4}{\|\bz\|^6}\left\|\frac{1}{m}\sum_{k=1}^m2(\ba_k^\top\bz)^4(\ba_k^\top\bx)^2h'\left(\frac{|\ba_k^\top\bz|^2}{\|\bz\|^2}\right)h\left(\frac{m|\ba_k^\top\bx|^2}{\|\by\|_1}\right)\bz\bz^\top \right\|\\
&+\frac{4}{\|\bz\|^6}\left\|\frac{1}{m}\sum_{k=1}^m(\ba_k^\top\bz)^2(\ba_k^\top\bx)^4h'\left(\frac{|\ba_k^\top\bz|^2}{\|\bz\|^2}\right)h\left(\frac{m|\ba_k^\top\bx|^2}{\|\by\|_1}\right)\bz\bz^\top \right\|\\
    &\lesssim \frac{4}{\|\bz\|^6}\cdot \|\bz\|^8\cdot |h'|_{\infty}\cdot \gamma^3\left(\sqrt{\beta}e^{-0.49\beta}+\epsilon\right)+\frac{8}{\|\bz\|^6}\cdot \|\bz\|^6\|\bx\|^2\cdot |h'|_{\infty}\cdot 2\gamma^3\left(\sqrt{\beta}e^{-0.49\beta}+\epsilon\right)\\
    &+\frac{4}{\|\bz\|^6}\cdot \|\bz\|^4\|\bx\|^4\cdot |h'|_{\infty}\cdot 4\gamma^3\left(\sqrt{\beta}e^{-0.49\beta}+\epsilon\right)\\
    &\lesssim\gamma^3\left(\sqrt{\beta}e^{-0.49\beta}+\epsilon\right)\cdot |h'|_{\infty}\max\left\{\|\bz\|^2,\|\bx\|^2,\frac{\|\bx\|^4}{\|\bz\|^2} \right\},
\end{align*}
where the third inequality follows from  the fact $\|\bz\bz^\top\|=\|\bz\|^2$ and Lemma \ref{lem:tech_lemma4} with $(s,t)=(6,0),(4,2),(2,4)$, respectively.

\paragraph{Bound for $B_{12}$}
\begin{align*}
B_{12}&\leq\frac{1}{\|\bz\|^4}\left|\frac{1}{m}\sum_k\left[ (\ba_k^\top\bz)^6-2(\ba_k^\top\bz)^4(\ba_k^\top\bx)^2+(\ba_k^\top\bz)^2(\ba_k^\top\bx)^4\right]h'\left(\frac{|\ba_k^\top\bz|^2}{\|\bz\|^2}\right)h\left(\frac{m|\ba_k^\top\bx|^2}{\|\by\|_1}\right) \right|\\
&\leq\frac{1}{\|\bz\|^4}\left|\frac{1}{m}\sum_k(\ba_k^\top\bz)^6h'\left(\frac{|\ba_k^\top\bz|^2}{\|\bz\|^2}\right)h\left(\frac{m|\ba_k^\top\bx|^2}{\|\by\|_1}\right) \right|\\
&+\frac{1}{\|\bz\|^4}\left|\frac{1}{m}\sum_k2(\ba_k^\top\bz)^4(\ba_k^\top\bx)^2h'\left(\frac{|\ba_k^\top\bz|^2}{\|\bz\|^2}\right)h\left(\frac{m|\ba_k^\top\bx|^2}{\|\by\|_1}\right) \right|\\
&+\frac{1}{\|\bz\|^4}\left|\frac{1}{m}\sum_k(\ba_k^\top\bz)^2(\ba_k^\top\bx)^4h'\left(\frac{|\ba_k^\top\bz|^2}{\|\bz\|^2}\right)h\left(\frac{m|\ba_k^\top\bx|^2}{\|\by\|_1}\right) \right|\\
    &\lesssim \frac{1}{\|\bz\|^4}\cdot \|\bz\|^6\cdot |h'|_{\infty}\cdot \gamma^3\left(\sqrt{\beta}e^{-0.49\beta}+\epsilon\right)+ \frac{2}{\|\bz\|^4}\cdot \|\bz\|^4\|\bx\|^2\cdot |h'|_{\infty}\cdot 2\gamma^3\left(\sqrt{\beta}e^{-0.49\beta}+\epsilon\right)\\
    &+ \frac{1}{\|\bz\|^4}\cdot \|\bz\|^2\|\bx\|^4\cdot |h'|_{\infty}\cdot 4\gamma^3\left(\sqrt{\beta}e^{-0.49\beta}+\epsilon\right)\\
    &\lesssim\gamma^3\left(\sqrt{\beta}e^{-0.49\beta}+\epsilon\right)\cdot |h'|_{\infty}\max\left\{\|\bz\|^2,\|\bx\|^2,\frac{\|\bx\|^4}{\|\bz\|^2} \right\} 
\end{align*}
where the third inequality follows from Lemma \ref{lem:tech_lemma4} with $(s,t)=(6,0),(4,2),(2,4)$, respectively.\\

Noting that $1<\beta<\gamma$, combing the above bounds together yields \eqref{eq:thmR1_eq2a0}.
\bibliographystyle{abbrv}
\bibliography{ref}
\end{document}